\documentclass[aps,prd,twocolumn,showpacs, superscriptaddress,preprintnumbers,amsmath,amssymb,floatfix]{revtex4}

\usepackage{amsmath,amssymb}
\usepackage[colorlinks]{hyperref}

\usepackage{color}
\usepackage[dvips]{graphicx}

\newcommand{\lsim}{\mathrel{\mathop{\kern 0pt \rlap
  {\raise.2ex\hbox{$<$}}}
  \lower.9ex\hbox{\kern-.190em $\sim$}}}
\newcommand{\gsim}{\mathrel{\mathop{\kern 0pt \rlap
  {\raise.2ex\hbox{$>$}}}
  \lower.9ex\hbox{\kern-.190em $\sim$}}}
\newcommand{\sigmav}{\langle \sigma_{\rm ann} v \rangle}

\newcommand{\beq}{\begin{equation}}
\newcommand{\eeq}{\end{equation}}
\newcommand{\bea}{\begin{eqnarray}}
\newcommand{\ena}{\end{eqnarray}}

\newcommand{\eVdist}{\kern-0.06667em}
\def\Mp{M_{\rm Pl}}
\def\vp{\varphi}
\def\beqra{\begin{eqnarray}}
\def\eeqra{\end{eqnarray}}

\def\beq{\begin{equation}}
\def\eeq{\end{equation}}

\def\A{\mathcal A}

\begin{document}

\preprint{DFTT 71/2009}

\title{Thermal Relics in Modified Cosmologies: Bounds on Evolution Histories
of the Early Universe and Cosmological Boosts for PAMELA}

\author{R. Catena} 
\email{r.catena@thphys.uni-heidelberg.de}
\affiliation{Institut f\"ur Theoretische Physik \\
Philosophenweg 16, D-69120, Heidelberg, Germany}

\author{N. Fornengo}
\email{fornengo@to.infn.it}
\affiliation{Dipartimento di Fisica Teorica, Universit\`a di Torino, via P. Giuria 1, I--10125 Torino, Italy}
\affiliation{Istituto Nazionale di Fisica Nucleare, via P. Giuria 1, I--10125 Torino, Italy}

\author{M. Pato}
\email{pato@iap.fr}
\affiliation{Dipartimento di Fisica, Universit\`a di Padova  \\ via
Marzolo 8, I-35131, Padova, Italy}
\affiliation{Institut d'Astrophysique de Paris \\ 98bis bd Arago, 75014, Paris, France}
\affiliation{Universit\'e Paris Diderot-Paris 7 \\ rue Alice Domon et L\'eonie Duquet 10, 75205, Paris, France}

\author{L. Pieri}
\email{lidia.pieri@gmail.com}
\affiliation{Dipartimento di Astronomia, Universit\`a di Padova  \\ 
vicolo dell'Osservatorio 3, I-35122, Padova, Italy}
\affiliation{Istituto Nazionale di Fisica Nucleare, Sezione di Padova \\ via
Marzolo 8, I-35131, Padova, Italy}

\author{A. Masiero}
\email{masiero@pd.infn.it}
\affiliation{Dipartimento di Fisica, Universit\`a di Padova  \\ via
Marzolo 8, I-35131, Padova, Italy}
\affiliation{Istituto Nazionale di Fisica Nucleare, Sezione di Padova \\ via
Marzolo 8, I-35131, Padova, Italy}

\date{\today}

\begin{abstract} 
Alternative cosmologies, based on extensions of General Relativity, predict modified thermal histories
in the Early Universe during the pre Big Bang Nucleosynthesis (BBN) era, epoch which is not directly constrained by cosmological observations.
When the expansion rate is enhanced with respect to the standard case, thermal relics typically decouple 
with larger relic abundances. The correct value of the relic abundance is therefore obtained for larger 
annihilation cross--sections, as compared to standard cosmology. A direct consequence is that indirect detection
rates are enhanced. Extending previous analyses of ours, we derive updated astrophysical bounds on the dark matter 
annihilation cross sections and use them to constrain alternative cosmologies in the pre--BBN
era. We also determine the characteristics of these alternative cosmologies in order to provide
the correct value of relic abundance for a thermal relic for the (large) annihilation
cross--section required to explain the PAMELA results on the positron fraction, therefore providing a 
``cosmological boost" solution to the dark matter interpretation of the PAMELA data. 
\end{abstract}

% 11.30.Pb Supersymmetry
% 12.60.Jv Supersymmetric models
% 95.30.Cq Elementary particle processes
% 95.35.+d Dark matter
% 98.35.Gi Galactic halo (Milky Way)
% 98.35.Df Kinematics, dynamics, and rotation
% 98.35.Pr Solar neighborhood

% 95.36.+x Dark energy
% 98.80.-k Cosmology
% 04.50.+h Gravity in more than four dimensions, KK theory, unified
%          field theories, alternatives theories of gravity
% 96.50.S- Cosmic rays in the solar system
% 98.70.Sa Cosmic rays
% 98.80.Cq Particle theory and field theory models in the early Universe

\maketitle

%%%%%%%%%%%%%%%%%%%%%%%%%%%%%%%%%%%%%%%%%%%%%%%%%%%%%%%%%%%%%%%%%%%%%%%%%

\section{Introduction}
\label{sec:intro}

Big Bang Nucleosynthesis (BBN) is the deepest available probe of the early Universe. Its success in explaining the primordial abundances of light elements \cite{PDG,Steigman}, combined with Cosmic Microwave Background (CMB) and Large Scale Structure (LSS) studies, confirms the standard model of cosmology since the BBN epoch at MeV temperatures. At those temperatures the universe must have been essentially radiation-dominated. Before, however, a period of very enhanced expansion may have occured. In two earlier works \cite{Schelke:2006eg,Donato:2006af} some of us have used indirect searches for Dark Matter (DM) annihilation, namely antiprotons and $\gamma$--rays from the Galactic Centre, to derive limits on the pre--BBN expansion rate of the universe. The basic idea is as follows. If DM is composed of Weakly Interacting Massive Particles (WIMPs) which thermalize in the Early Universe and then freeze--out their
abundance before BBN, then the expansion history since freeze--out and the precise WMAP measurement of the dark matter relic abundance $(\Omega h^2)_{CDM}^{WMAP}=0.1131 \pm 0.0034$ \cite{WMAP5y} fix the annihilation cross-section $\langle \sigma_{ann}v \rangle$ (for a given DM mass $m_{DM}$). A faster pre--BBN expansion requires a larger annihilation cross--section in order to meet the relic abundance bound, and  in turn enhanced DM--induced astrophysical fluxes result. Thus it is possible to draw an upper limit on the Hubble rate $H(T)$ before the BBN epoch \cite{Schelke:2006eg,Donato:2006af}. In the present work we revisit this subject mainly motivated by the host of astrophysical data released in the last years, such as cosmic--ray electrons and positrons (PAMELA \cite{pamelae}, ATIC \cite{atic2}, Fermi-LAT \cite{fermi}, HESS \cite{hess08,hess09}), antiprotons (PAMELA \cite{pamelapbar}) and $\gamma$-rays (HESS \cite{hessgc}, Fermi-LAT \cite{Fermiinter,Fermidiff}). The rising behaviour of the positron fraction observed by PAMELA \cite{pamelae}, in particular, has been posed under deep scrutiny, and in addition to astrophysical interpretations 
\cite{pulsars,Blasi,BlasiSerpico} it has been discussed the possibility that the raise is due to DM annihilation
dominantly occurring into leptons \cite{ArkaniHamed}. The DM interpretation
requires large values of $\sigmav$, which are then incompatible with a successfull 
thermal relic. Mechanisms have been put forward in order to solve this problem \cite{ArkaniHamed,zentner}.
Furthermore, results from N--body simulations have been recently presented \cite{VL2,Aq} allowing for a better modelling of dark structure and substructure in our Galaxy. 

Consequently, we are now in a  position to reassess 
the constraints on pre--BBN cosmologies by using a rather complete scheme of observables. Besides an update of the previous works \cite{Schelke:2006eg,Donato:2006af}, we are also interested in the possibility of accomodating the rising positron fraction with annihilations of a thermal DM particle whose properties naturally arise in non--standard cosmologies. Refs.~\cite{Schelke:2006eg,Donato:2006af,Catena:2004ba,nonstd1,nonstd2}, for instance, show that the large annihilation cross-sections required to meet the measured positron fraction or electron spectrum are attainable in the context of non--standard cosmological scenarios such as low reheating temperatures, scalar--tensor theories of Gravity, kination phases or brane world cosmology. Analyses of phenomenological consequences of modified cosmologies are also discussed
in Ref.s \cite{alternative}.

In this paper we pursue a generic parameterisation of the pre--BBN expansion rate \cite{Schelke:2006eg,Donato:2006af} and consider a multi-messenger, multi-wavelength scheme of DM constraints. We then close with an explicit simple
realization of the cosmological enhancement mechanism in a scalar--tensor theory of Gravity.

%%%%%%%%%%%%%%%%%%%%%%%%%%%%%%%%%%%%%%%%%%%%%%%%%%%%%%%%%%%%%%%%%%%%%%%%%
\section{Survey of bounds on annihilating Dark Matter}
\label{sec:survey}

\begin{figure}[htp]
\includegraphics[width=1.0\columnwidth]{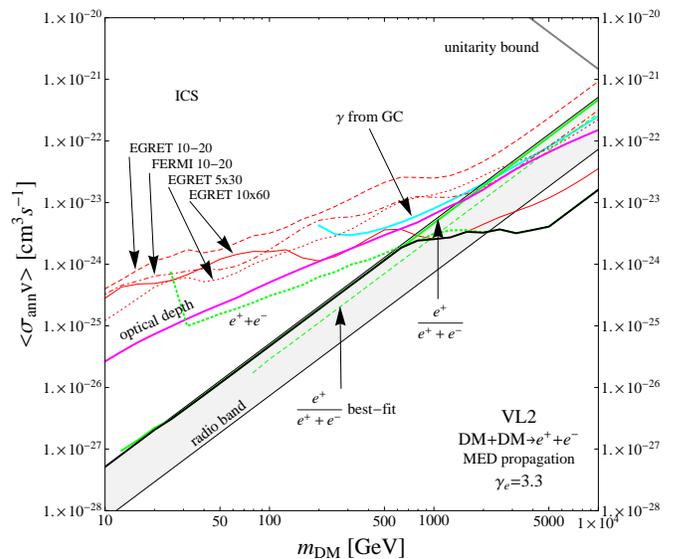}
\caption{Upper limits on the DM annihilation cross--section (versus the DM mass) coming from different astrophysical observations. Here we have considered the Via Lactea II DM distribution, DM annihilations into $e^+e^-$, the MED propagation model for cosmic rays and an electron spectral index $\gamma_e=3.3$. The region above the thick black line is excluded by the convolution of all the implemented constraints. The shaded band labelled as ``radio band'' denotes the uncertainty 
on the radio constraint. The dashed line labelled as ``$\frac{e^+}{e^++e^-}$ best--fit'' denotes the
values of the DM annihilation cross--section required to explain the PAMELA data on the positron fraction. The unitarity bound assuming s-wave annihilations \cite{beacom} is also shown.}
\label{fig:boundsVL2ee.eps}
\end{figure}

\begin{figure}[htp]
\includegraphics[width=\columnwidth]{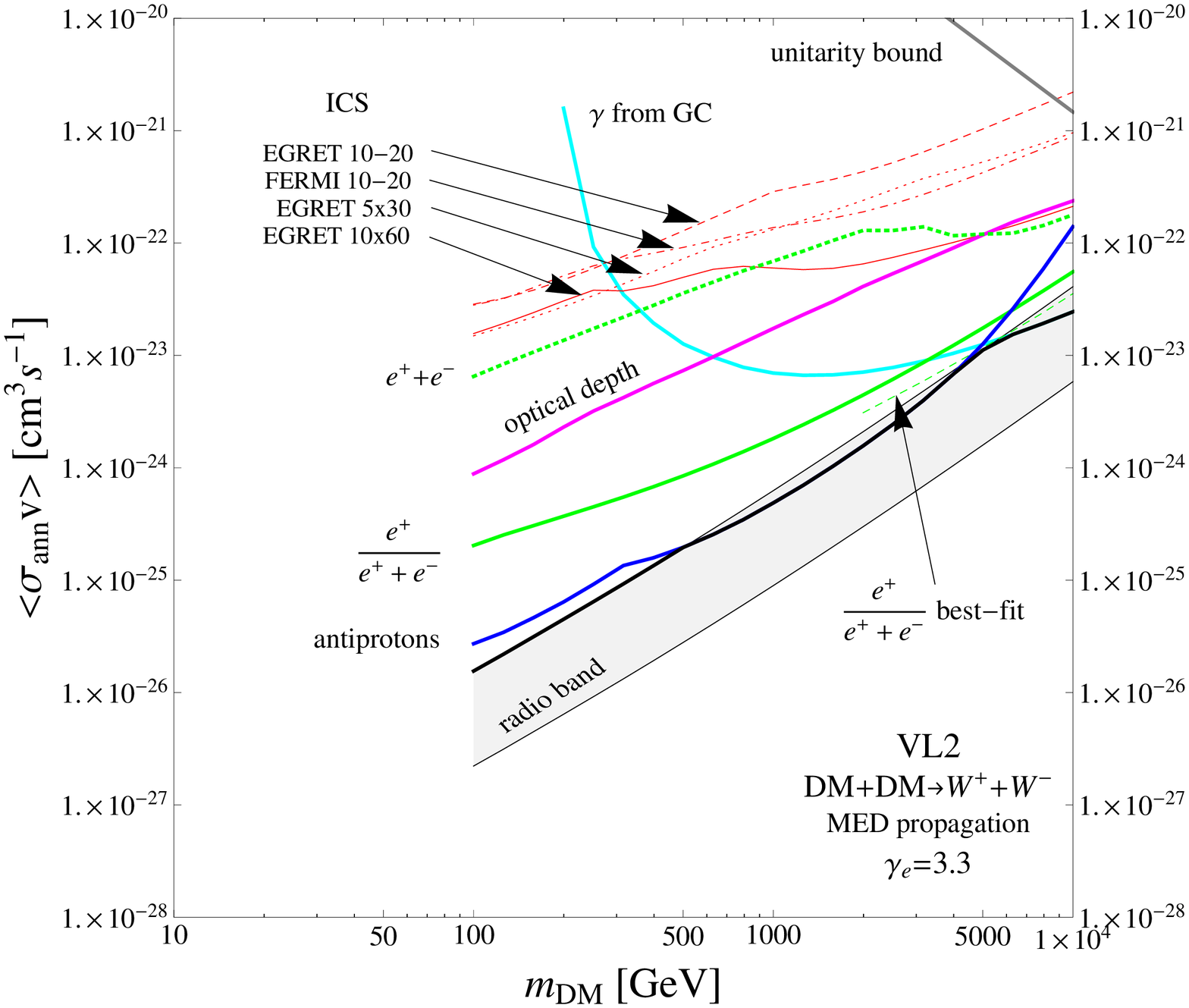}
\caption{The same as in Fig.~\ref{fig:boundsVL2ee.eps}, for DM annihilations into $W^+W^-$.}
\label{fig:boundsVL2WW.eps}
\end{figure}

\begin{figure}[htp]
\includegraphics[width=\columnwidth]{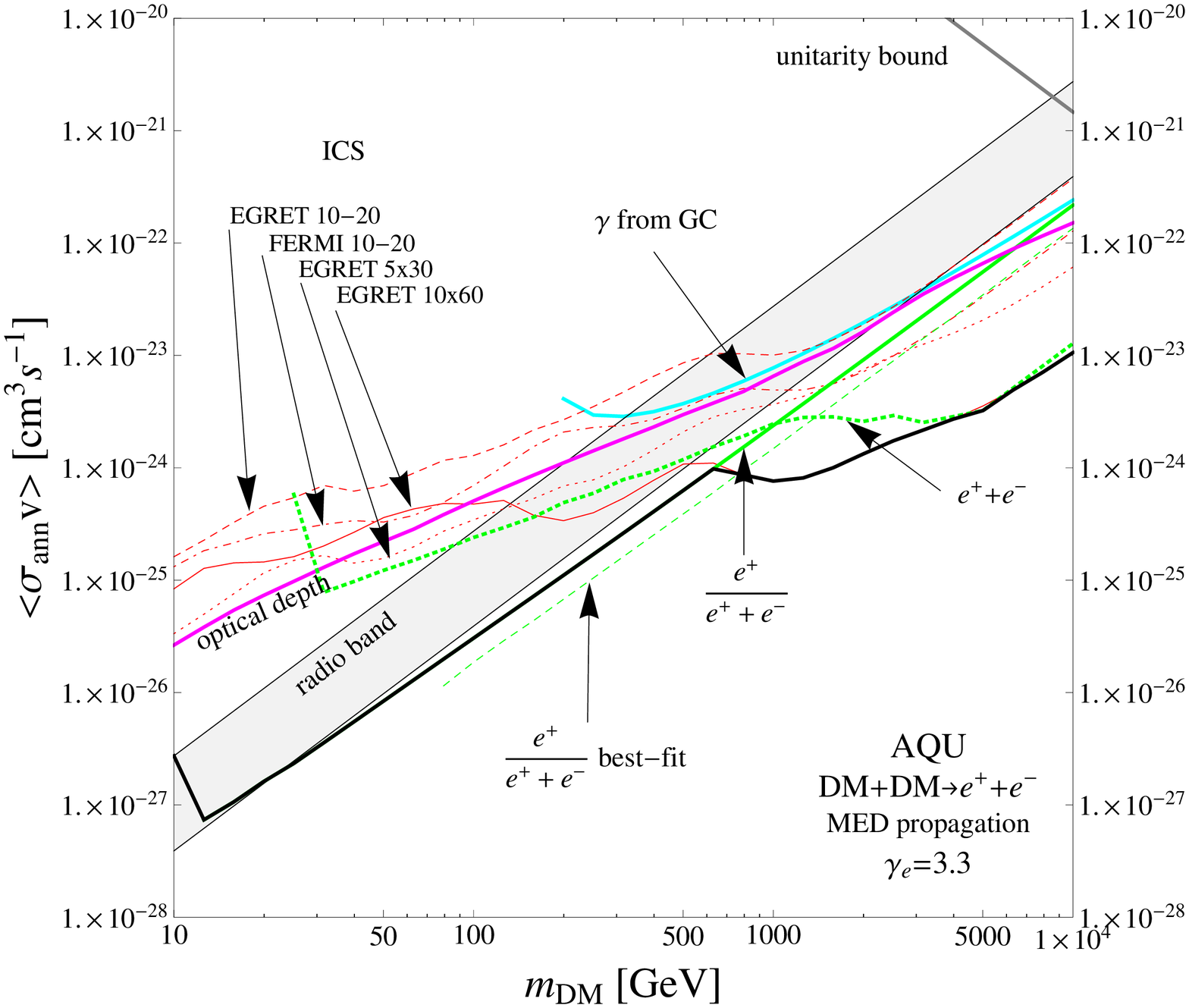}
\caption{The same as in Fig.~\ref{fig:boundsVL2ee.eps}, for Aquarius DM distribution.}
\label{fig:boundsAQUee.eps}
\end{figure}

\begin{figure}[htp]
\includegraphics[width=\columnwidth]{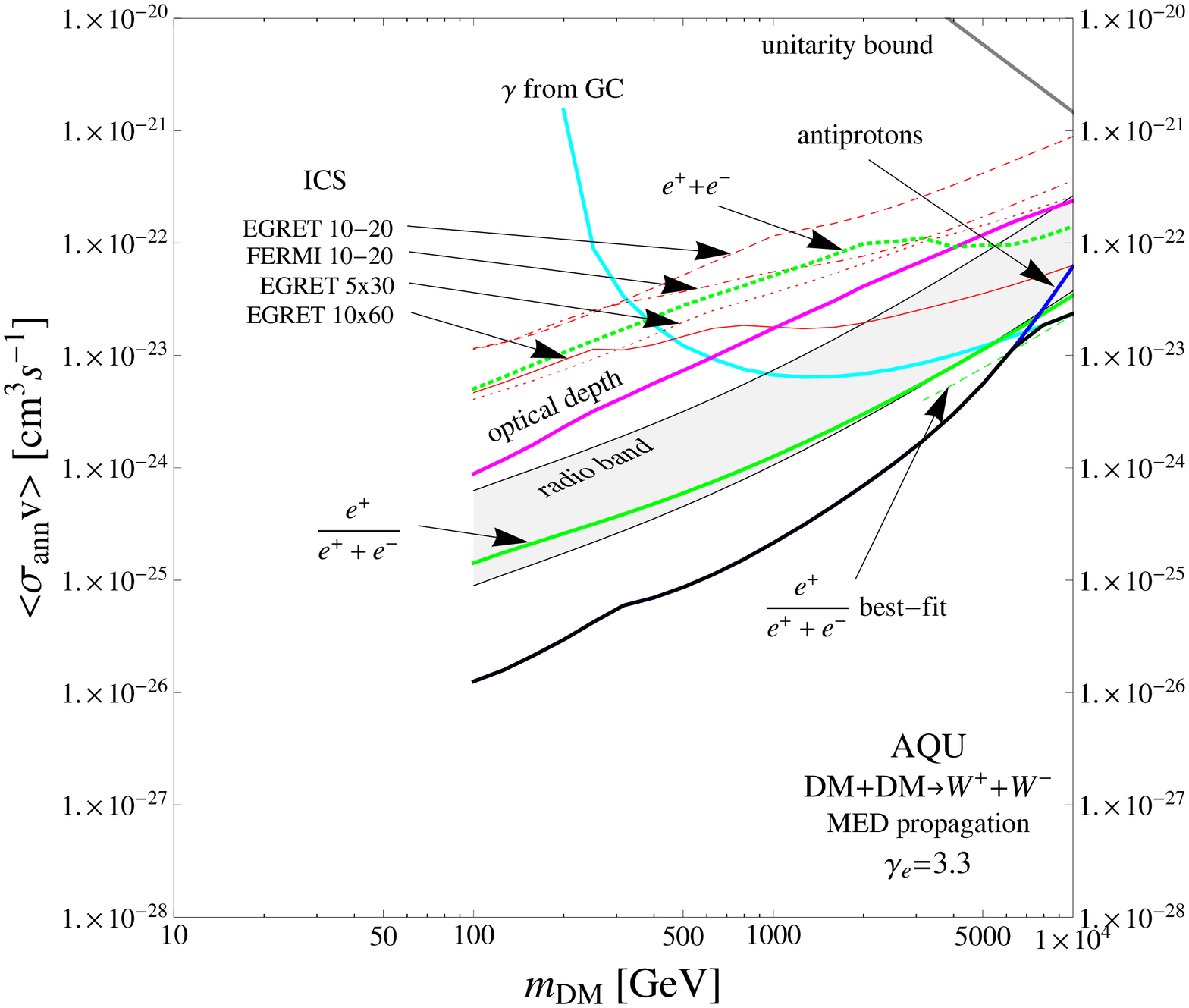}
\caption{The same as in Fig.~\ref{fig:boundsVL2WW.eps}, for Aquarius DM distribution.}
\label{fig:boundsAQUWW.eps}
\end{figure}

\begin{figure}[htp]
\includegraphics[width=\columnwidth]{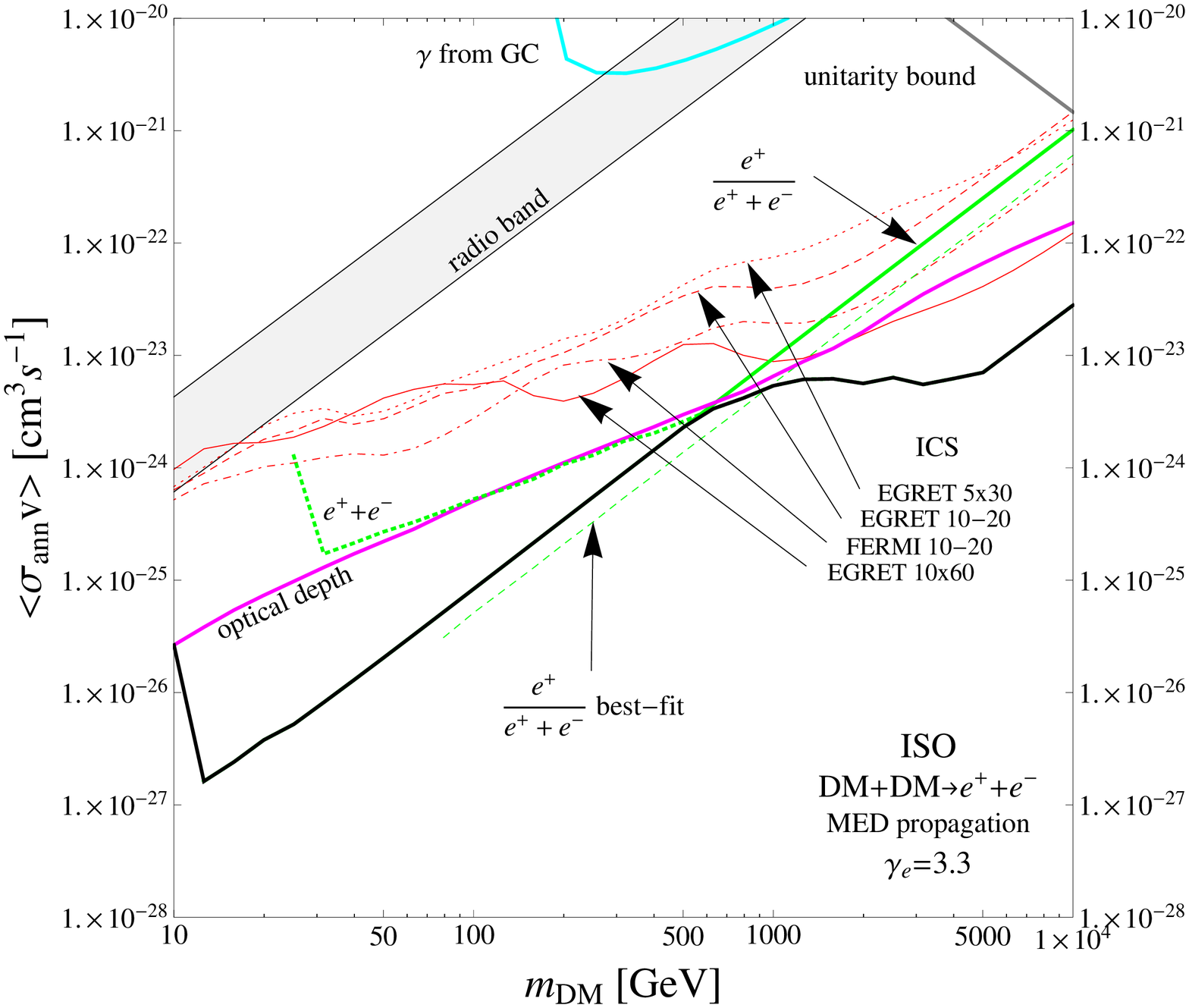}
\caption{The same as in Fig.~\ref{fig:boundsVL2ee.eps}, for a cored isothermal DM distribution.}
\label{fig:boundsISOee.eps}
\end{figure}

\begin{figure}[htp]
\includegraphics[width=\columnwidth]{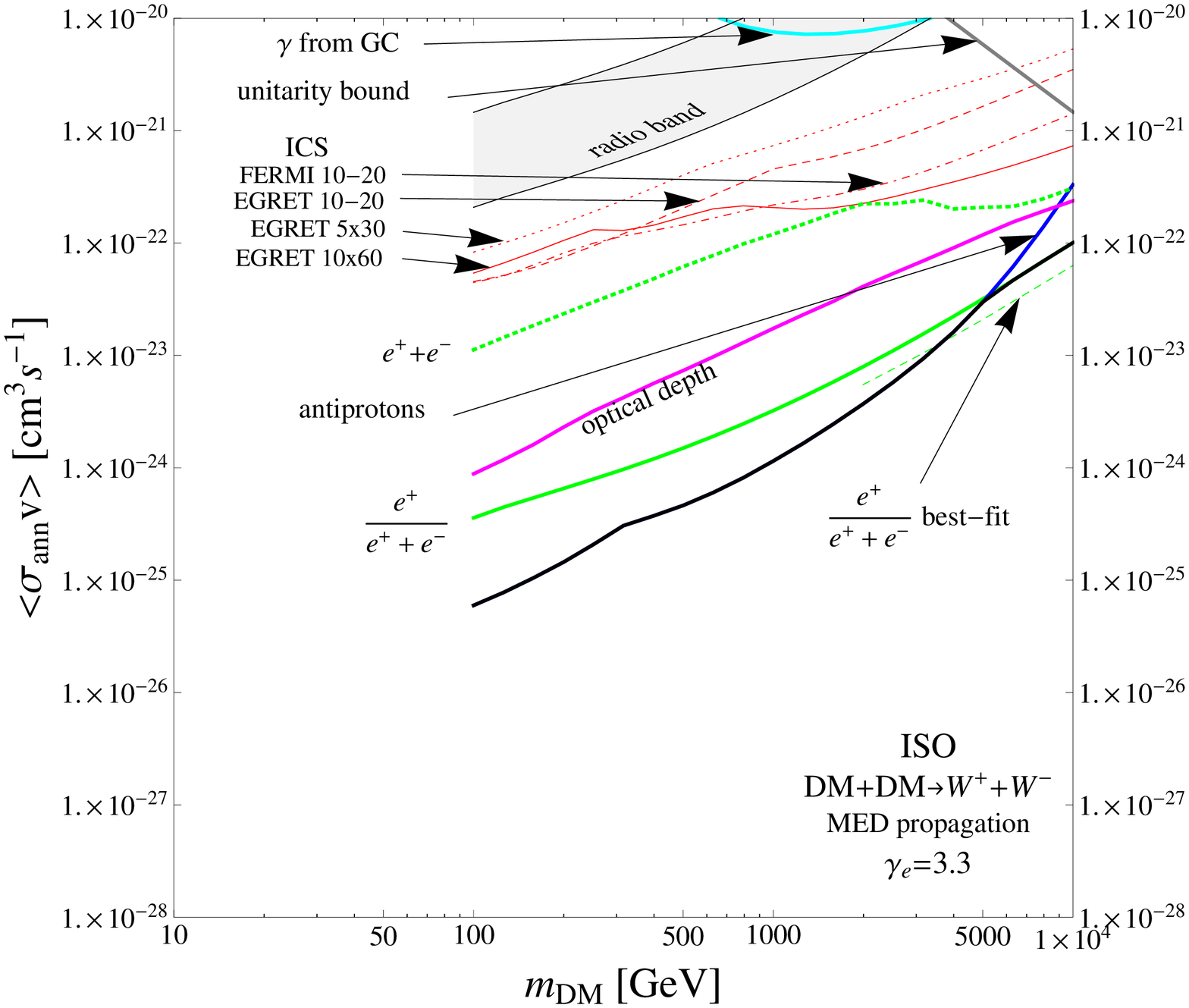}
\caption{The same as in Fig.~\ref{fig:boundsVL2WW.eps}, for a cored isothermal DM distribution.}
\label{fig:boundsISOWW.eps}
\end{figure}

\begin{figure}[htp]
\includegraphics[width=\columnwidth]{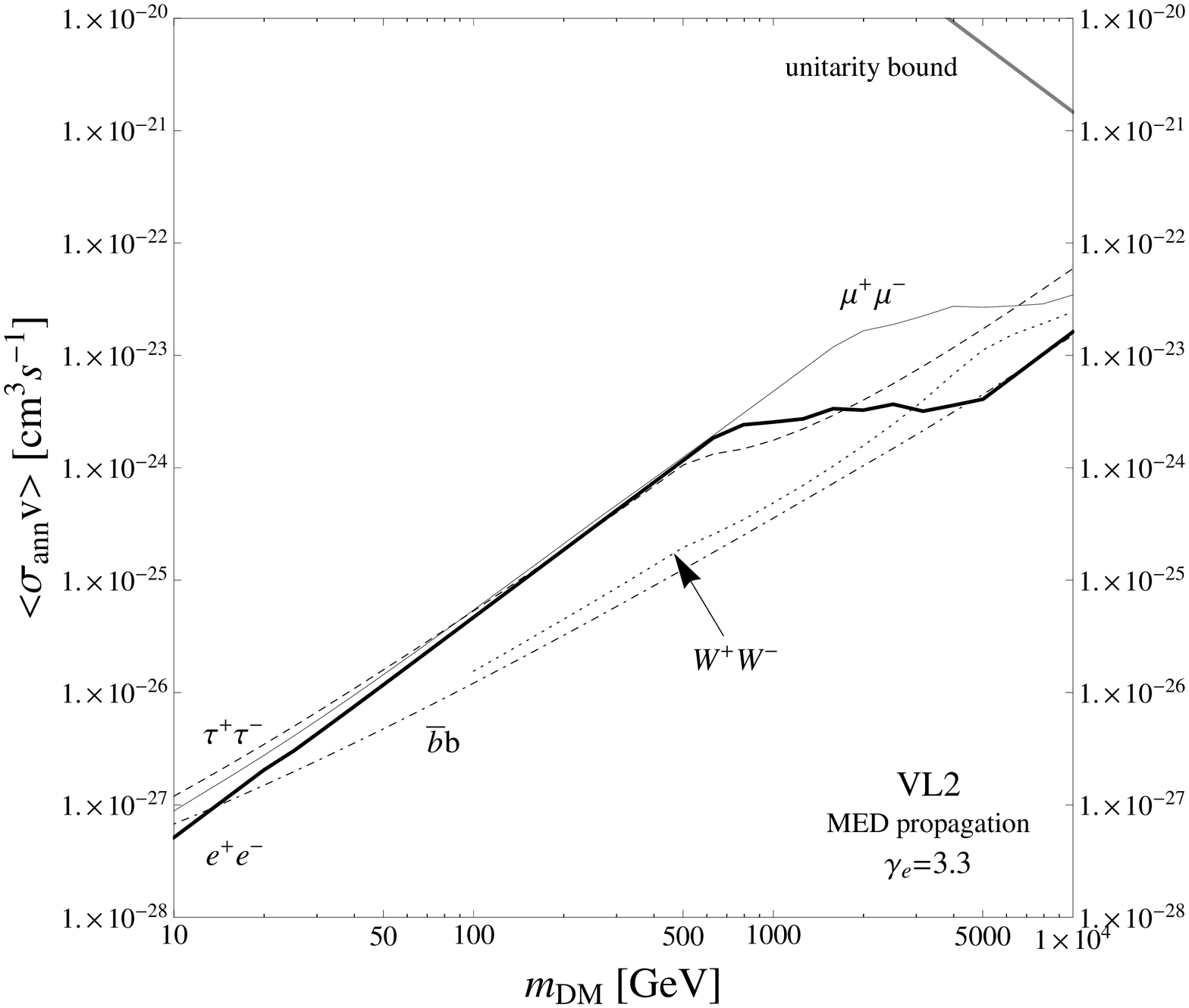}
\caption{Summary of the astrophysical bounds on the DM annihilation cross--section vs. the DM mass, for the Via Lactea II DM distribution and for different DM annihilation channels: $e^+e^-$, $\mu^+\mu^-$, $\tau^+\tau^-$, $W^+W^-$ and $\bar{b}b$. Cosmic rays are propagated in the MED model and the electron spectral index is $\gamma_e=3.3$.}
\label{fig:boundsVL2all.eps}
\end{figure}

\begin{figure}[htp]
\includegraphics[width=\columnwidth]{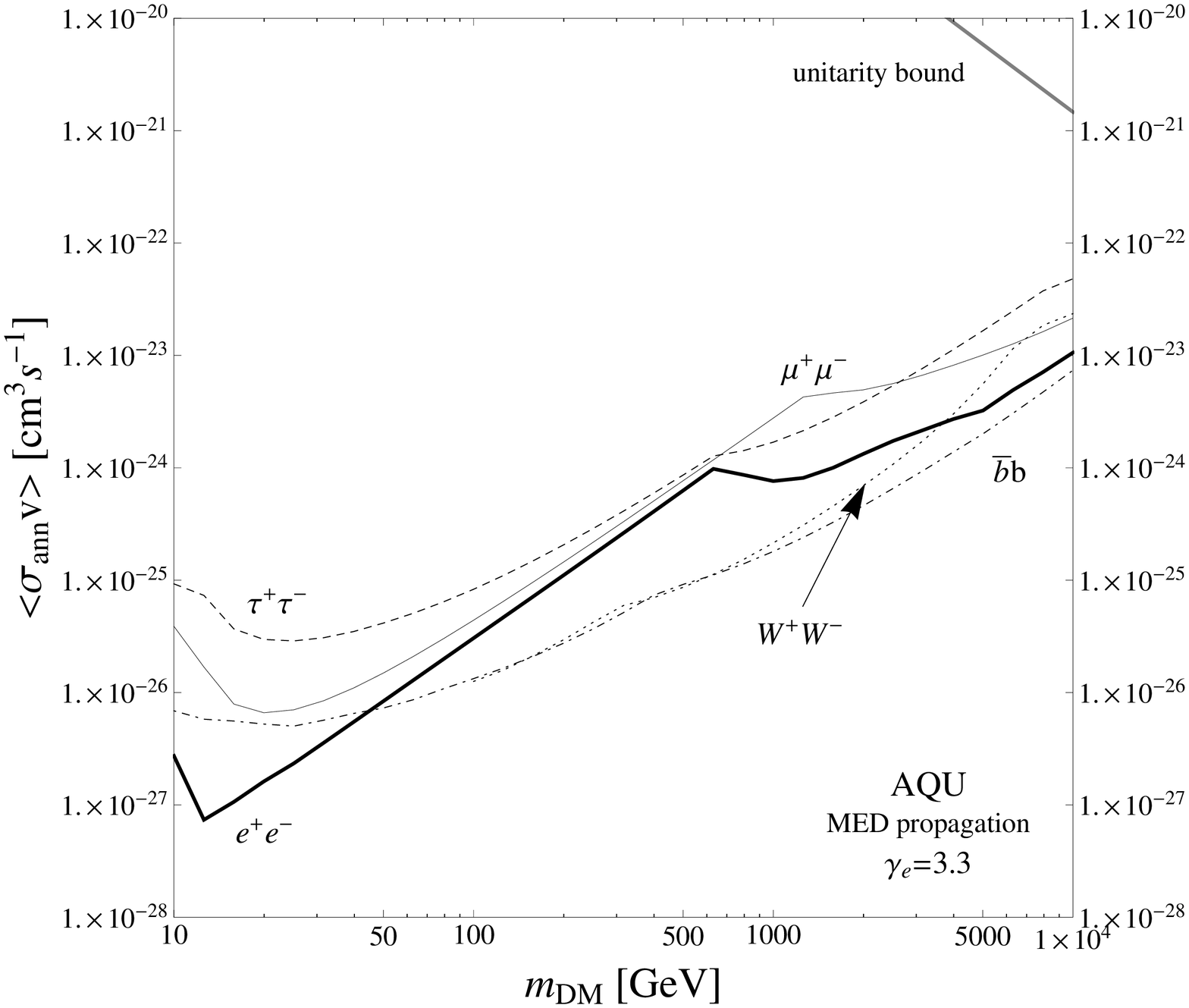}
\caption{Summary of the astrophysical bounds on the DM annihilation cross--section vs. the DM mass, for the Aquarius DM distribution and for different DM annihilation channels: $e^+e^-$, $\mu^+\mu^-$, $\tau^+\tau^-$, $W^+W^-$ and $\bar{b}b$. Cosmic rays are propagated in the MED model and the electron spectral index is $\gamma_e=3.3$.}
\label{fig:boundsAQUall.eps}
\end{figure}

\begin{figure}[htp]
\centering
\includegraphics[width=\columnwidth]{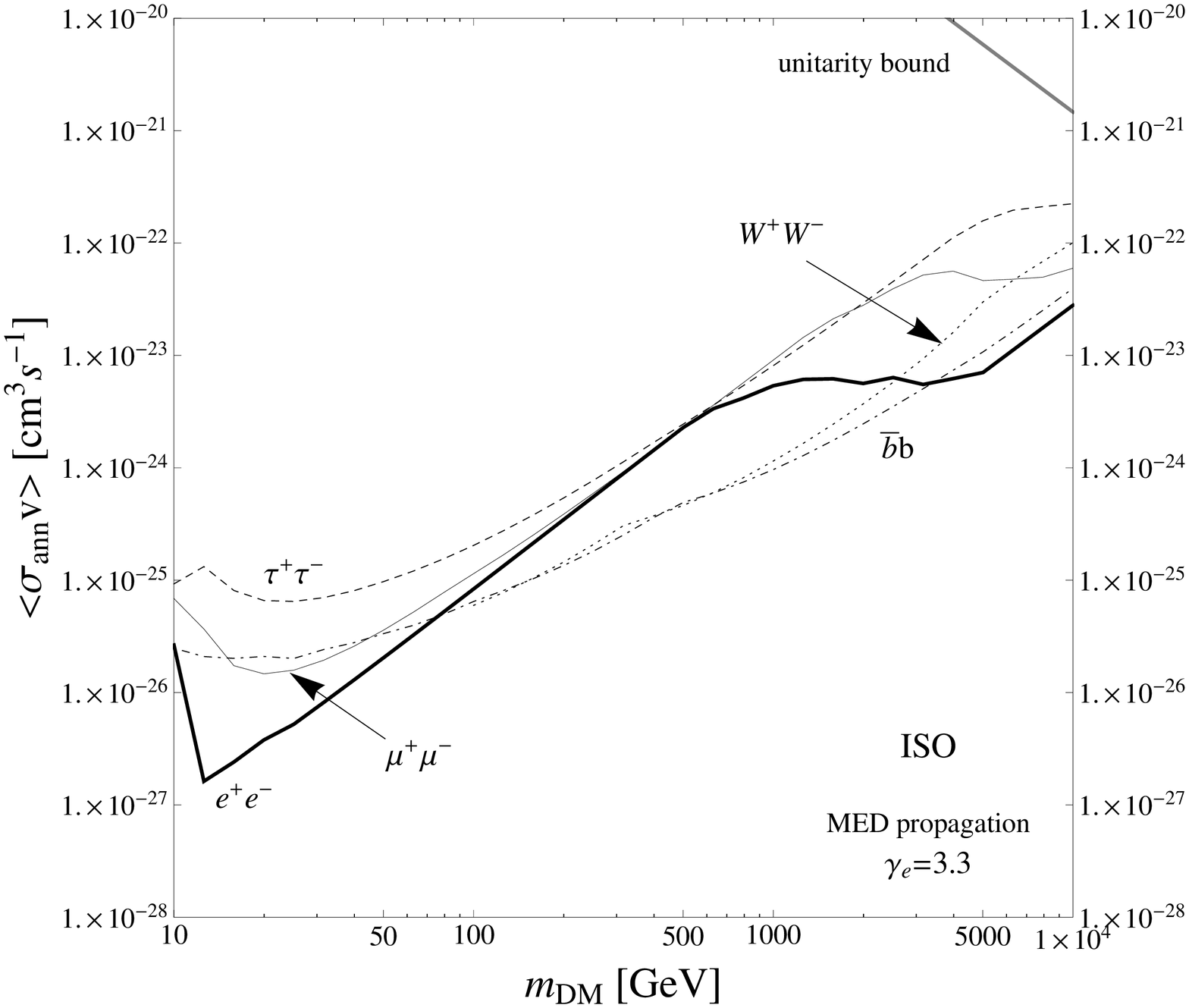}
\caption{Summary of the astrophysical bounds on the DM annihilation cross--section vs. the DM mass, for a cored isothermal DM distribution and for different DM annihilation channels: $e^+e^-$, $\mu^+\mu^-$, $\tau^+\tau^-$, $W^+W^-$ and $\bar{b}b$. Cosmic rays are propagated in the MED model and the electron spectral index is $\gamma_e=3.3$.}
\label{fig:boundsISOall.eps}
\end{figure}

Dark Matter annihilations could be sources of electrons and positrons, (anti)protons, (anti)deuterons, photons and (anti)neutrinos. Such particles, or their interactions in the surrounding medium, provide a plethora of ways to effectively constrain the intrinsic properties of annihilating DM (for a review see e.g.~\cite{BHS}). Here we will focus on galactic positrons, antiprotons, $\gamma$-rays and radio photons, and on constraints related to the optical depth of Cosmic Microwave Background photons. Other relevant channels, which are not included in our analysis, are neutrinos from the Galactic Centre (GC) \cite{nuGC} (neutrinos from Sun \cite{nuSun,Bere} and Earth \cite{Bere} do not
directly constrain the total annihilation cross section), and $\gamma$-rays from dwarf spheroidal galaxies \cite{gadSph,Bertone1}, galaxy clusters \cite{gaclus} and extragalactic halos \cite{gaextra}. For multi-messenger, multi-wavelength analyses see e.g.~\cite{Bertone1,Regis,Donato08,Bergstrom1,PPB}.

In the present work we adopt a model--independent approach and consider generic annihilating DM candidates of masses $m_{DM}$ in the interval $\left[10\textrm{ GeV},10\textrm{ TeV}\right]$. We study the bounds 
on the velocity--averaged annihilation cross-section $\sigmav$ for each annihilation
channel separately: $e^+e^-$, $\mu^+\mu^-$, $\tau^+\tau^-$, $W^+W^-$ or $\bar{b}b$. This scheme basically captures the essential features of several well--motivated DM particles such as the lightest supersymmetric particle (LSP) or the lightest Kaluza-Klein particle (LKP). We will consider throughout the paper the case of a DM particle which
dominantly annihilates as s--wave.

Annihilation signals are proportional to the square of the dark matter density, and it is therefore crucial to properly specify the DM distribution in the Milky Way. We rely on the recent results of the high--resolution N-body simulations Via Lactea II (VL2) \cite{VL2} and Aquarius (Aq) \cite{Aq}. In the former, the total DM profile (smooth + clumpy components) is well fitted to a Navarro-Frenk-White (NFW) profile, while in the latter the density distribution follows an Einasto profile with $\alpha$ = 0.17. Once the clumpy component has been subtracted to the virial mass of the simulated Milky--Way--like halos, the local density of the smooth component is $\rho_{\odot}^{VL2}=0.41\textrm{ GeV/cm}^3$ and $\rho_{\odot}^{Aq}=0.58\textrm{ GeV/cm}^3$. Extrapolating the results of the simulations down to a minimal subhalo mass for WIMPs of $10^{-6}\textrm{ M}_{\odot}$ (e.g.~\cite{Green}), one gets $\sim 53(17)$\% of the total Milky Way mass distributed in virialised substructures for Via Lactea II (Aquarius) (see Ref.~\cite{LPBB} for further details and analysis of the findings of the two simulations). Notice that different WIMPs and different early universe expansion rates may induce minimal subhalo masses significantly smaller or larger than the fiducial value $10^{-6}\textrm{ M}_{\odot}$. However, this would not change significantly our results or conclusions. In fact, as shown by \cite{LPBB}, the clump fraction becomes important at about 70(100) kpc for Via Lactea II (Aquarius) simulations. This is because tidal forces deplete the presence of subhalos in the central part of the halo. Therefore, the radio, $\gamma$-ray from the galactic centre (and to less extent, electron and positron) constraints depend only on regions where the subhalo component is not dominant. The inverse Compton scattering constraints, which could depend in principle on the subhalo abundances, turn out to be unimportant in our analysis, while the CMB constraints refer to a period where no structures had formed thus being insensitive the problem. As an additional benchmark for the galactic distribution of DM, we consider the case of a smooth isothermal profile with no substructures. Following \cite{Fornengo1,CatenaUllio}, we set $\rho_{\odot}^{\rm iso}=0.4\textrm{ GeV/cm}^3$.

The astrophysical bounds on the dark matter annihilation cross section $\sigmav$, as a function
of the dark matter mass $m_{DM}$, for the different annihilation channels and for the different
dark matter density profiles, are summarized in Figs.~\ref{fig:boundsVL2ee.eps}--\ref{fig:boundsISOall.eps}. 
Due to the instrinsic complexity of these figures, we refer the reader to the internal labellings and
to the figure captions in order to pin down the various bounds, without excessive comments in the body
of the article.
We just remark that Figs.~\ref{fig:boundsVL2ee.eps} to \ref{fig:boundsISOWW.eps} show the separate impact of all the astrophysical bounds,
for specific annihilation final states and specific DM halo profiles. Figs. \ref{fig:boundsVL2all.eps}, 
\ref{fig:boundsAQUall.eps} and \ref{fig:boundsISOall.eps}
represent a summary of the astrophysical bounds on the DM annihilation cross--section vs.~the DM mass, for the Via Lactea II,
Aquarius and cored isothermal halo profiles, respectively. We notice that the effect of the bounds may be non
trivial, depending on the DM mass, the annihilation channels and the halo profile: the most stringent bound arises as
a combination of different observational sources, it is tipically a factor of 10 (up to 100) stronger for the
Via Lactea II DM distribution, as compared to the isothermal profile (with Aquarius typically staying in between).
Clearly, signals which are more sensitive to the galactic center DM shape (like the radio bound or gamma--rays from
the galactic center) induce bounds which are more affected by the uncertainty on the DM distribution.
This is clearly seen in the radio band constraint in Fig.s \ref{fig:boundsVL2ee.eps}, 
\ref{fig:boundsAQUee.eps} and \ref{fig:boundsISOee.eps} for the $e^+e^-$ channel
and Fig.s \ref{fig:boundsVL2WW.eps}, \ref{fig:boundsAQUWW.eps} and \ref{fig:boundsISOWW.eps} for the $W^+W^-$ channel: moving from Via Lactea II to Aquarius, the bound
loosens by about one order of magnitude; the bound is instead weakened by about 4 orders of magnitude for an
isothermal DM distribution. Other signals, like e.g.~antiprotons, are not very sensitive to the DM halo
profile \cite{Donato:2003xg}, as can be seen by confronting again 
Fig.s \ref{fig:boundsVL2WW.eps}, \ref{fig:boundsAQUWW.eps} and \ref{fig:boundsISOWW.eps}. Instead, signals like antiprotons or positrons
are more sensitive to the diffusive region of the galactic environment \cite{Donato:2003xg,Delahaye07}, and therefore
will produce bounds which are limited by the uncertainties on the cosmic--rays propagation properties: the effect
of this will be shown in Section \ref{sec:analysis}.

\subsection{Antimatter}

Unlike neutral particles, positrons and antiprotons produced in the Milky Way undergo different processes that change their direction and energy while crossing the galactic medium. The galactic magnetic fields, for instance, are responsible for deflection and, due to their (poorly known) inhomogeneities, the evolution of a positron or an antiproton can be treated as a random walk with a certain diffusion coefficient. For positrons, important phenomena to take into account are energy losses through inverse Compton scattering off the CMB and starlight and synchrotron emission, which proceed at a space--independent rate $b(E_{e^{+}})\simeq E_{e^{+}}^2/(\textrm{GeV}\cdot\tau_E)$ with $\tau_E\simeq 10^{16}$ s \cite{Delahaye08,Delahaye07} and result in a diffusion length of a few kpc. Antiprotons, on the other hand, can travel larger distances without losing much energy by synchrotron or inverse Compton since $m_p\gg m_e$. Instead, they may be swept away by galactic winds, assumed to be constant and perpendicular to the disk. Furthermore, annihilations $p\bar{p}$ mainly in the disk are responsible for the disappearance of primary antiprotons. To compute propagated fluxes we apply the formalism outlined in \cite{Donato:2003xg,lavalle} (and references therein) and use the MIN, MED and MAX propagation models. In this framework, the total positron or antiproton flux at Earth for a specific dark matter candidate and a certain propagation model is the sum of the smooth and clumpy DM components:
\begin{equation}
\phi^{\rm DM}(E)=(1-f_{\odot})^2 \, \phi_{\rm sm} (E) \, + \, \langle \phi_{\rm cl} \rangle (E) \, ,
\end{equation}
where $f_{\odot}=3.0(0.23)\%$ is the local clump fraction for VL2 (Aq) and $E$ plays the role of kinetic energy in the antiproton case. Recall that both $\phi_{\rm sm}$ and $\langle \phi_{\rm cl} \rangle$ are proportional to $\sigmav$. We notice that the term $\langle \phi_{\rm cl} \rangle$ has been modelled following \cite{lavalle} and folds self-consistently the population of subhalos and the Green functions for the propagated positrons/antiprotons.

\subsubsection{Electrons and positrons}

The PAMELA satellite has measured the positron fraction $\phi_{e^+}/(\phi_{e^+}+\phi_{e^-})$ in the energy range 1.5$-$100 GeV \cite{pamelae}. The data show a steep rise above $\sim 7$ GeV. Here we restrict ourselves to $E_{e^+} > 10$ GeV and therefore disregard solar modulation. In order to translate the PAMELA positron data into an upper limit on the DM-induced positron flux $\phi_{e^+}^{\rm DM}$, we assume (\emph{i}) a power-law electron flux $\phi_{e^-}\propto E_{e^-}^{-\gamma_e}$, $\gamma_e=\left\{3.3,3.4,3.5\right\}$ (in rough agreement with \cite{Casadei}) normalised to the AMS--01 measurement at 10 GeV \cite{AMS01}; and (\emph{ii}) a secondary positron flux $\phi_{e^+}^{sec}$ produced by the spallation of cosmic rays in the Galaxy \cite{Delahaye08}. In this scheme, $\phi_{e^+}=\phi_{e^+}^{DM}+\phi_{e^+}^{\rm sec}$. We now require, for fixed $m_{DM}$, that no single energy bin of the PAMELA positron data above 10 GeV is exceeded by more than 3$\sigma$. This produces an overall upper limit on $\sigmav$. A complementary bound, particularly for multi-TeV leptophilic DM candidates, is motivated by the recent measurements of the electron plus positron flux by Fermi-LAT \cite{fermi} and HESS \cite{hess08,hess09}. As with PAMELA positron fraction, we disregard solar modulation -- Fermi-LAT and HESS cover energies ranging from $\sim$ 25 to $\sim$ 5000 GeV -- and, conservatively, draw the 2$\sigma$ upper limit on $\sigmav$ by only considering electrons and positrons produced by DM annihilations.

On the other hand, we are also interested in studying the feasibility of explaining the PAMELA positron fraction with DM annihilations. So, for fixed $m_{DM}$, we fit the data above 10 GeV leaving $\langle \sigma_{ann}v \rangle$ as a free parameter and demand that $\chi^2<20$ (7 data points are available). Whenever the positron best--fit values of $\sigmav$ are not in
conflict with bounds coming from other astrophysical signals, we consider that value as a positive solution of
the PAMELA data in terms of dark matter annihilation. More detailed fitting procedures have been presented in other references e.g.~\cite{Cirelli1}. It is well--known that it is possible to obtain good fits to the PAMELA positron data with DM particles which annihilate preferably into leptons (hadrons) with masses above $\sim 100\textrm{ GeV}$ (a few TeV). Of course, the required annihilation cross-sections are larger than the standard thermal value, $2.1\cdot10^{-26}\textrm{ cm}^3\textrm{s}^{-1}$. We clearly recover these results in our analysis. Differences with respect to Refs.~such as \cite{Cirelli1} are due to a different local dark matter density (we are using values greater than the ``usual" 0.3 GeV/cm$^3$), the inclusion of dark matter substructure (according to Via Lactea II and Aquarius simulations), and the use of a slighlty smaller electron flux (we normalise to AMS-01 at 10 GeV and consider a power-law index 3.3 as a reference value). All such factors play to lower our ``PAMELA best-fit" cross-sections with respect to other Refs.

\par Note as well that a joint explanation of the PAMELA positron fraction and the Fermi-LAT/HESS electron plus positron flux in terms of DM annihilations favours heavy rather than light DM particles. However, in this work we do not pursue a global fit to PAMELA/Fermi-LAT/HESS but instead use the electron plus positron flux as an astrophysical constraint only.

\subsubsection{Antiprotons}

The comparison of the PAMELA antiproton ratio $\phi_{\bar{p}}/\phi_{p}$ \cite{pamelapbar} with theoretical estimates of secondary antiprotons reveals little space for $\bar{p}$ deriving from DM annihilations (or any other primary source) \cite{Donato08}. These considerations disfavour light DM particles decaying prominently into hadrons.

To derive the antiproton bound we use the interstellar proton flux $\phi_{p}$ and the interstellar secondary antiproton flux $\phi_{\bar{p}}^{\rm sec}$ as given in Ref.~\cite{Donato08}, and apply a solar modulation in the force field approximation with $\phi_F=500$ MV. The 2$\sigma$ upper bound on $\sigmav$ from antiproton searches is derived by
using the whole energy spectrum in the range where PAMELA antiproton measurements are available \cite{pamelapbar}.

\subsection{$\gamma$-rays}

In some specific models, DM particles can annihilate directly into photon(s) producing clear spectral $\gamma$-ray lines (see e.g.~\cite{Gustafsson}): the branching ratios for such processes are usually rather low. We do not consider in this paper such annihilation channels. Instead, we consider $\gamma$-rays produced by decays of (or radiation from) final state particles of the annihilation process. These processes lead to a continuous spectrum up to energies close to the mass of the DM particle (e.g.~\cite{Bergstrom1997,Ullio2002}):
\begin{equation}\label{promptgammas}
\frac{d \Phi_\gamma}{dE_\gamma} =
 \frac{1}{4 \pi} \frac{\sigmav}{2 m^2_{DM}} \cdot \frac{d N_\gamma}{d E_\gamma} \int_{V}  \frac{\rho_{DM}^2}{d^2} dV \, ,
\end{equation}
where $d N_\gamma/d E_\gamma$ is the differential spectrum per annihilation for a given annihilation channel \cite{FPS04,Bergstrom1} (we remind that we assume all the annihilation proceeds via a single channel). We compute the $\gamma$-ray flux in a $10^{-5}$ sr solid angle towards the Galactic Centre (GC). We compare our predictions with the HESS measurement of the GC source in 2003 and 2004, $\Phi (> 160\textrm{ GeV}) = 1.89 \pm 0.38 \times 10^{-11} \textrm{ }\gamma\textrm{ cm}^{-2} \textrm{s}^{-1}$ \cite{hessgc}, and derive the bound on the annihilation cross-section requiring that our model does not exceed the measurement by more than 2$\sigma$: this is the bound from the 
galactic center, labelled as ``$\gamma$ from GC'' in Figs.~\ref{fig:boundsVL2ee.eps}--\ref{fig:boundsISOWW.eps}.

Another way for DM annihilations to give rise to $\gamma$--ray fluxes is through inverse Compton scattering (ICS)
on electrons and positrons produced by DM annihilation. In fact, low--energy photons, such as those in the CMB, starlight and infrared radiation, may be up--scattered by 
high--energy electrons and positrons. This channel has gained particular relevance ever since the PAMELA collaboration published their exciting results: if the positron data are due to DM annihilations, then there should exist a large population of electrons and positrons in the Milky Way able to up--scatter low--energy photons. In Ref.~\cite{Panci} the authors computed the ICS $\gamma$--ray spectrum towards regions far from the GC. Here we apply the same procedure. Notice that these constraints are especially robust against the DM density profile since the GC region is excluded from the field of view.

\subsection{Radio photons}

The interpretation of the rising positron fraction observed by PAMELA as a signal of DM annihilations would result in a large amount of highly energetic electrons and positrons permeating our Galaxy. This should hold true particularly towards the GC where the DM density is expected to be the highest. Such relativistic electrons and positrons propagating in the galactic magnetic field emit synchrotron radiation in the radio frequency band. Let us focus on a region towards the GC, small enough so that diffusion does not play an important role and where the galactic magnetic field is strong enough to neglect electron (and positron) energy losses other than synchrotron emission. Assuming further that advection is negligible and following Refs.~\cite{Bertone1,Bergstrom1}, the total synchrotron power emitted by the distribution of DM--induced electrons and positrons is:
\begin{equation}\label{nudwdnu}
\nu \frac{d W_{\rm syn}}{d\nu}= \frac{\langle \sigma_{ann} v \rangle}{2 m_{DM}^2} \int_{V_{\rm obs}}{dV \, \rho_{DM}^2 (\textbf{x}) \, E_{p}(\textbf{x},\nu) \, \frac{N_{e^{\pm}}(>E_p)}{2}  } \, ,
\end{equation} 
where $E_p(\textbf{x},\nu)=\sqrt{4\pi m_e^3 \nu/(3\cdot 0.29 e B(\textbf{x}))}$. We compute solely the contribution given by the smooth distribution of DM. Following Ref.~\cite{Bertone2}, we consider a cone with half-aperture 4'' pointed toward the GC and $\nu=0.408$ GHz for which an upper bound of 0.05 Jy has been derived from radio observations \cite{radioobs}. In this way we can determine the constraint on the plane $m_{DM}-\sigmav$ given by radio observations of the GC. In order to take into account the different approaches of Refs.~\cite{Regis} and \cite{Bertone1}, we weaken the bound obtained with Eq.~\eqref{nudwdnu} by a factor 7 \footnote{We thank Marco Regis and Marco Taoso for the numerical comparison between Refs.~\cite{Regis,Bertone1}.} and consider such rescaled result as an effective constraint. The radio bound is shown as a shaded band in Figs.~\ref{fig:boundsVL2ee.eps}--\ref{fig:boundsISOWW.eps}.

\subsection{Optical depth of CMB photons}

The optical depth of CMB photons depends on how and when the re--ionization of the universe occured. If DM is annihilating, then a considerable amount of high--energy electrons and positrons may be created after recombination giving rise, by ICS on CMB photons, to a population of (low--energy) $\gamma$--rays. These can easily ionize the gas releasing electrons and hence reducing the optical depth of CMB photons \cite{Hooper1}. Comparing such effect with the measured optical depth by WMAP, the authors of Ref.~\cite{CIP09} have derived upper limits on the annihilation cross--section of DM particles, that we also reproduce in Figs.~\ref{fig:boundsVL2ee.eps}--\ref{fig:boundsISOWW.eps}. Since there is clearly no dependence on the DM density profile, these constraints are very robust and difficult to avoid. This constraint has also been carefully analyzed in Refs. \cite{Slatyer:2009yq} and 
\cite{Huetsi:2009ex}.

Other effects of conspicuous DM annihilations in the early Universe are the heating of the intergalactic medium \cite{CIP09} and the distortion of CMB anisotropies and polarisation \cite{Galli}.

%%%%%%%%%%%%%%%%%%%%%%%%%%%%%%%%%%%%%%%%%%%%%%%%%%%%%%%%%%%%%%%%%%%%%%%%%

\section{Modified cosmologies}
\label{sec:alternative}

Cosmological models arising in modification of General Relativity (GR) very often 
predict a cosmological history with an expansion rate $H(T)$ 
larger, at early times, than the Hubble expansion rate $H_\textrm{GR}(T)$ 
of standard cosmology. Generically, we can encode this enhancement into
a temperature--dependent function $A(T)$ as \cite{Catena:2004ba,Schelke:2006eg,Donato:2006af}:
\begin{eqnarray}
H(T)&=&A(T)H_{\textrm{{GR}}}(T)
\label{eq:modhubble}
\end{eqnarray}
with $A(T) > 1$ at large temperatures and with $A(T) \rightarrow 1$ before BBN
sets up, in order not to spoil the successful predictions of BBN on the abundance
of primordial light elements.

With an increased expansion rate, thermal relics freeze--out their abundance earlier
than in standard cosmology: this implies that a thermal (cold) relic 
%realizes the so--called ``WIMP miracle'' 
matches the correct relic abundance
for annihilation cross--sections $\sigmav$ which are larger than in
standard cosmology. A consequence of this is that dark matter particles possess today,
in the galactic environment, larger annihilation cross--sections and thence enhanced
indirect detection signals, as compared to those obtained for a thermal decoupling
in GR. This implies that indirect searches for dark matter, like those discussed in
the previous Section, may have a potential of
constraining pre--BBN cosmological histories, under the assumption that dark matter
is a thermal relic.

We have investigated the consequences of these phenomena in Refs.~\cite{Schelke:2006eg,Donato:2006af}, where
we used the cosmic--ray antiproton and gamma--ray data to derive bounds on the
admissible enhancement of the expansion rate in the pre--BBN phase. In the current paper
we extend these analyses to comprehend all the astrophysical observables dicussed in the
previous Section. 

In the case of positrons, the recent measurements from PAMELA on the positron fraction have
shown a steady ``anomalous'' rise at energies above 10 GeV, up to about 100 GeV (the current
largest probed energy). The interpretation 
of this so--called PAMELA excess is currently under deep investigation. Astrophysical origins
of this behavior have been shown to be able to explain the PAMELA data: local sources, like
pulsars \cite{pulsars}, or positron production mechanisms occurring inside the sources of cosmic--rays
\cite{Blasi,BlasiSerpico}
are suitable to reproduce quite well the PAMELA result. An alternative solution is offered
by dark matter annihilation: in this case, it has been clearly shown in many independent analyses
that in order to explain the PAMELA data, the dark matter candidate needs to meet a number
of requirements: first of all, the size of the annihilation rate has to be orders of magnitude larger than
the one obtained for a thermal relic in standard cosmology {\em i.e.}~referring to an annihilation
cross--section of $\sigmav = 2.1\cdot 10^{-26}$ cm$^3$ s$^{-1}$, which is the one required to obtain the correct relic abundance for cold DM $(\Omega h^{2})^{\rm WMAP}_{\rm CDM} = 0.11$ in standard cosmology; secondly
it has to dominantly annihilate into leptons, unless it is quite heavy 
(above the TeV scale), in this latter case it is allowed to decay also into hadronic channels, where
antiprotons are produced. Since alternative cosmologies with $A(T)>1$ imply that the correct
relic abundance of a relic particle is obtained with larger annihilaton cross-sections, they
offer a framework to explain the PAMELA data without requiring specific mechanisms to boost
the annihilation rate like, for  instance, Sommerfeld enhancements \cite{ArkaniHamed,zentner} or (unlikely) large 
astrophysical boosts \cite{lavalle}. In a sense, alternative cosmologies offer a ``cosmological boost''
to a thermal relic. In the first part of the next Section we will discuss under what conditions
PAMELA data are explained by means of this cosmological boost.

In order to be as general as possible, we will perform our analysis by parameterizing
the temperature--behavior of the enhancement function $A(T)$ as:

\begin{equation}
A(T)=1+\eta\left(\frac{T}{T_\textrm{f}}\right)^\nu
  \tanh\left(
 \frac{T-T_{\textrm{re}}}{T_{\textrm{re}}}\right)
\label{eq:Afunction}
\end{equation}
for temperatures $T>T_\textrm{re}$ and $A(T) = 1$ for $T \leq T_\textrm{re}$.
This form has been adopted in our previous analyses \cite{Catena:2004ba,Schelke:2006eg}:
it is a suitable parameterization to describe a cosmology where 
$H\rightarrow H_\textrm{GR}$, at some ``re-entering" temperature
$T_\textrm{re}$. 
We must require $T_\textrm{re}\gsim 1 \, \textrm{MeV}$ to make 
sure not to be in conflict with the predictions of BBN. For definiteness we
will fix $T_{\rm re} = 1$ MeV in our analysis, except when explicitely mentioned otherwise.
Notice that a sensitivity on this parameter is expected, as discussed
in Ref.~\cite{Catena:2004ba} and as it will be shown in the next Section. The maximal enchancement on the relic abundance is obtained for the lowest possible value of $T_{\rm re} = T_{\rm BBN}$. By fixing this parameter at 1 MeV, we derive the lowest bounds on the enhacement parameter $\eta$: larger $\eta$ are expected for larger $T_{\rm re}$.

For $T \gg T_\textrm{re}$ we have:
\begin{equation}
 A(T) = 1 + \eta \left(\frac{T}{T_\textrm{f}}\right)^\nu \longrightarrow \eta \left(\frac{T}{T_\textrm{f}}\right)^\nu 
\end{equation}
where the last implication is valid for $\eta \gg 1$. Thus, $T_\textrm{f}$ is the normalization temperature at which $A(T_\textrm{f}) = 1 + \eta$, which, again for large values of $\eta$, goes to $A(T_\textrm{f}) \longrightarrow \eta$. For definiteness, and to conform to our previous analyses, we take $T_\textrm{f}$ to be the temperature at 
which the WIMP DM candidate freezes out in standard cosmology ($T_\textrm{f}$ is therefore DM mass--dependent). Therefore
$1+\eta$ represents the enhancement of the Hubble rate at the time of the WIMP freeze--out. The freeze--out
temperature is determined with the standard procedure, which can be found for instance in 
Refs.~\cite{Catena:2004ba,Schelke:2006eg}.

We will organize our discussion in terms of bounds on $\eta$ for different cosmological models, characterized by the temperature--evolutionary parameter $\nu$: $\nu = 2$ refers to the Hubble rate evolution in a Randall--Sundrum type II brane cosmology scenario of Ref.~\cite{Randall};  
$\nu = 1$ is the typical kination evolution, discussed {\em e.g.} in Ref.~\cite{kination};
$\nu = -1$ is representative of the behavior found in scalar--tensor cosmologies in
Ref.~\cite{Catena:2004ba}, to which we refer for additional discussions on thermal relics in cosmologies with enhanced Hubble rate. The trivial case $\nu=0$ refers to an overall boost of the Hubble rate, like in the case of a large number of additional relativistic degrees of freedom in the thermal plasma.

In Section \ref{secModel} we will instead present a specific cosmological model arising in scalar--tensor
theories of gravity, where we will explicitely show the ability of such models to provide the
right amount of cosmological boost to explain the PAMELA results.

For more details on the modified cosmological scenarios, the calculation of
the relic abundance in these models, including some analytical results
and discussion, we refer to 
Refs.~\cite{Catena:2004ba,Schelke:2006eg,Donato:2006af}.

%%%%%%%%%%%%%%%%%%%%%%%%%%%%%%%%%%%%%%%%%%%%%%%%%%%%%%%%%%%%%%%%%%%%%%%%%

\section{Cosmological boost: PAMELA and bounds on modified cosmologies}
\label{sec:analysis}

\begin{figure}[t]
\includegraphics[width=1.1\columnwidth]{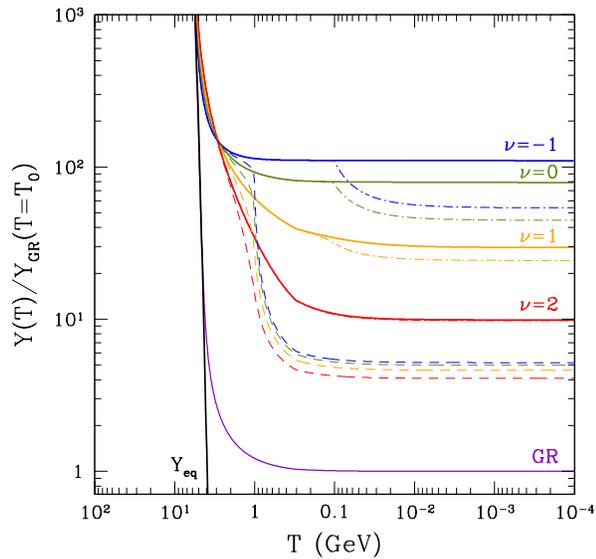}
\caption{Temperature evolution of the comoving abundance $Y$. The almost--vertical solid line shows the equilibrium
abundance, while the lowest solid line which asymptotizes to the relic abundance value refers to the
solution of the Boltzmann equation in standard cosmology (GR). The other solid lines refer to the solutions of the Boltzmann equation for different cosmologies with $T_{\rm re} = 1$ MeV: from top to bottom $\nu=-1,0,1,2$. 
Dot--dashed lines refer to $T_{\rm re} = 100$ MeV and dashed lines to $T_{\rm re} = 1$ GeV.
 All the lines are normalized to the asymptotic
value of the abundance in standard cosmology. The mass of the dark matter particle and annihilation cross section have been
fixed at: $m_\chi = 100$ GeV and $\sigmav = 2.1\cdot 10^{-26}$ cm$^3$ s$^{-1}$.}
\label{fig:decoupling.ps}
\end{figure}

\begin{figure}[t]
\includegraphics[width=1.1\columnwidth]{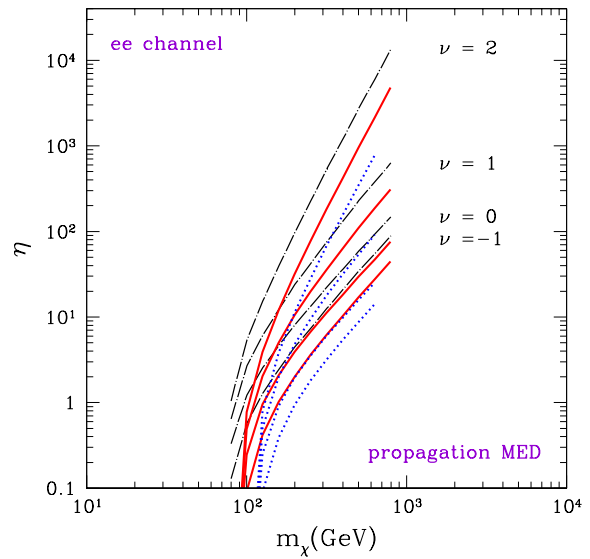}
\caption{Values of the $\eta$ parameter required to explain the PAMELA data together with $\Omega_{\chi} h^{2} = 
(\Omega h^{2})^{\rm WMAP}_{\rm CDM}$, in the case of DM annihilation into $e^{+}e^{-}$ and for different cosmologies labelled by the values of $\nu$. The DM annihilation cross section is required to explain the PAMELA data without violating the astrophysical bounds (see Figs.~\ref{fig:boundsVL2ee.eps}, \ref{fig:boundsAQUee.eps} and \ref{fig:boundsISOee.eps} for this case) and the values of $\eta$ are determined, for each DM mass, in order to have the correct DM relic abundance in the modified cosmology, which therefore produces the required cosmological boost. The solid (red) lines refer to the Via Lactea II DM distribution, the dot--dashed (black) lines refer to a cored isothermal sphere and the dotted (blue) lines refer to the Aquarius DM distribution. Propagation parameters are set at the MED case.}
\label{fig:PAMELA_eta_mchi_ee.ps}
\end{figure}

\begin{figure}[t]
\includegraphics[width=1.1\columnwidth]{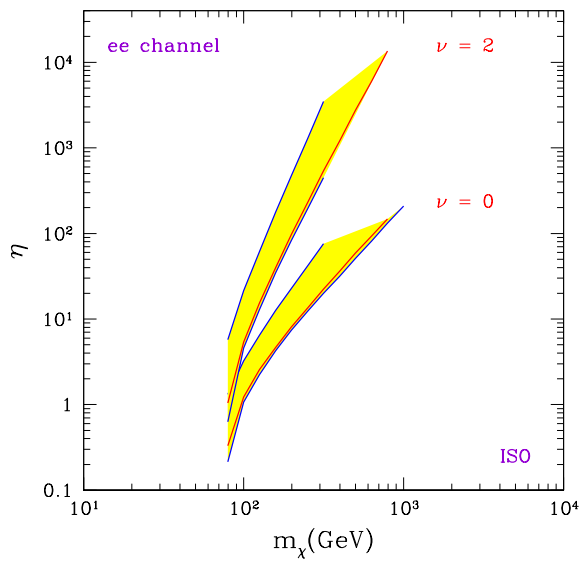}
\caption{The same as in Fig.~\ref{fig:PAMELA_eta_mchi_ee.ps}, but showing the uncertainty band coming from cosmic--ray propagation parameters: for each case, the upper line refers to the MIN model, the lower line to the MAX model. Only
cosmologies with $\nu=0$ and $\nu=2$ are shown. The DM distribution is a cored isothermal sphere.}
\label{fig:PAMELA_eta_mchi_ee_band.ps}
\end{figure}

\begin{figure}[t]
\includegraphics[width=1.1\columnwidth]{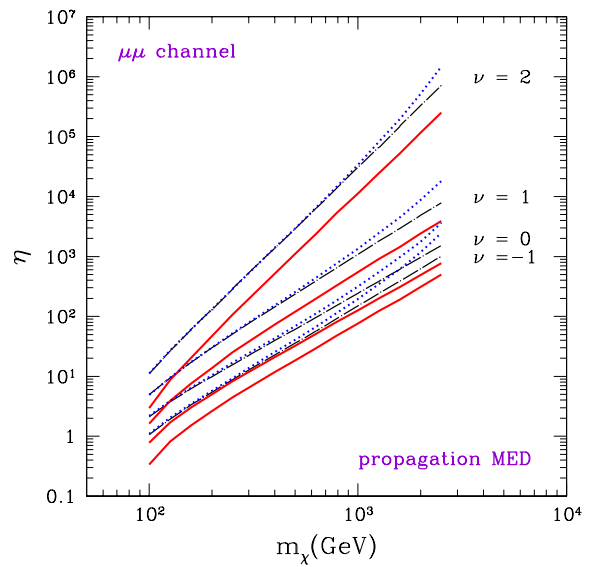}
\caption{The same as in Fig.~\ref{fig:PAMELA_eta_mchi_ee.ps}, for DM annihilation into $\mu^+\mu^-$ (except for the Aquarius case, not shown here).
The dotted lines refer instead to the case of $T_{\rm re} = 100$ MeV, for the cored isothermal dark matter distribution.}
\label{fig:PAMELA_eta_mchi_mumu.ps}
\end{figure}

\begin{figure}[t]
\includegraphics[width=1.1\columnwidth]{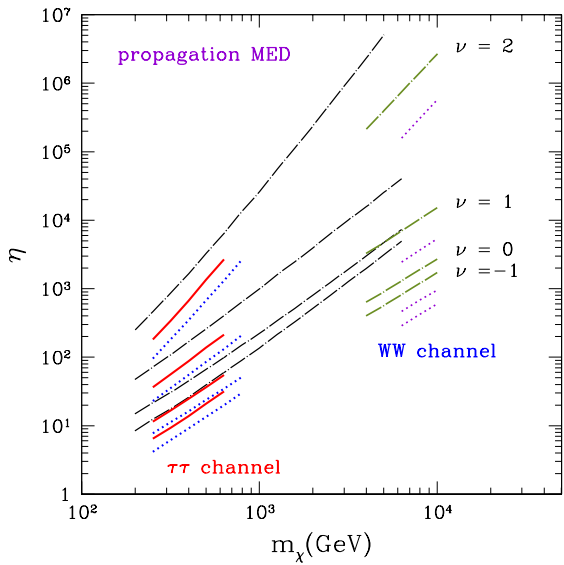}
\caption{The same as in Fig.~\ref{fig:PAMELA_eta_mchi_ee.ps}, for DM annihilation into $\tau^+\tau^-$ and into $W^+W^-$.}
\label{fig:PAMELA_eta_mchi_tautauWW.ps}
\end{figure}

\begin{figure}[t]
\includegraphics[width=1.1\columnwidth]{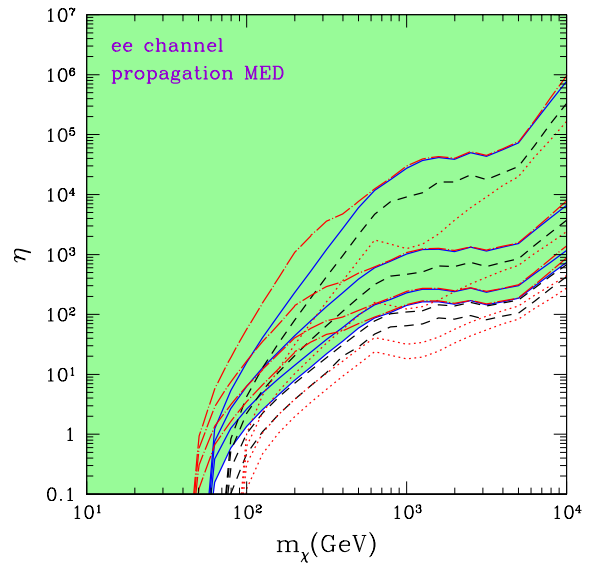}
\caption{Upper bound on the $\eta$ parameter for different cosmological models and for the case of DM annihilation into $e^+e^-$. The bounds arise from the astrophysical constraints on the DM annihilation cross--section and from the
requirement that the DM relic abundance matches the WMAP value for CDM. The solid (blue) lines refer to a cored DM distribution, the dashed (black) lines to the Via Lactea II DM distribution and the dotted (red) lines
to the Aquarius DM profile. Propagation parameters are set at the MED case.
The set of solid, dashed and dotted lines refers to cosmologies with $\nu=-1,0,1,2$ going from the lower to the upper curves.
For each cosmology, the excluded values of $\eta$ are those above the corresponding line. The red dot--dashed lines refer to a cored isothermal DM distribution and the MIN set of propagation parameters.}
\label{fig:BOUND_eta_mchi_ee.ps}
\end{figure}

\begin{figure}[t]
\includegraphics[width=1.1\columnwidth]{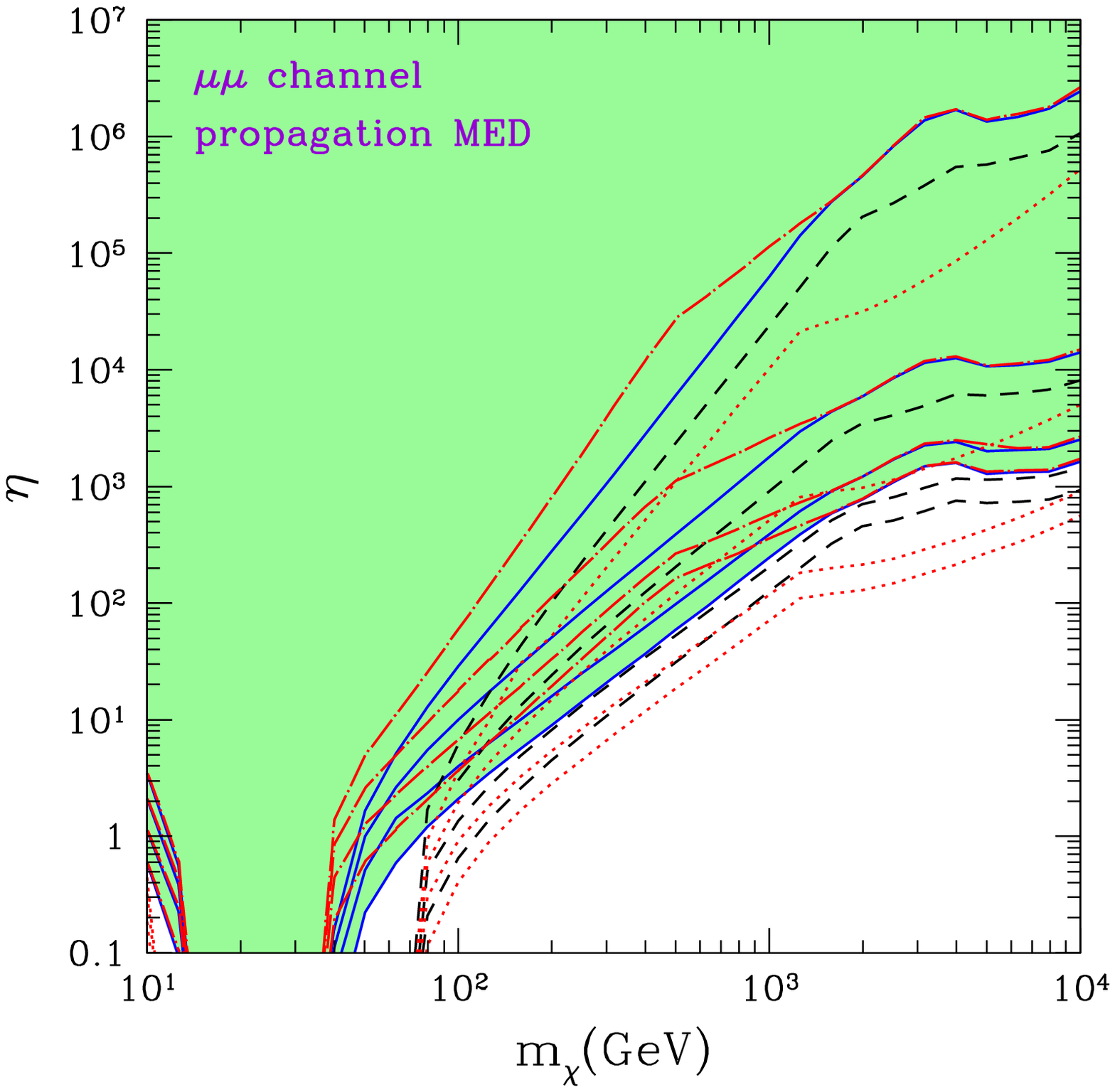}
\caption{The same as in Fig.~\ref{fig:BOUND_eta_mchi_ee.ps}, for DM annihilation into $\mu^+\mu^-$.}
\label{fig:BOUND_eta_mchi_mumu.ps}
\end{figure}

\begin{figure}[t]
\includegraphics[width=1.1\columnwidth]{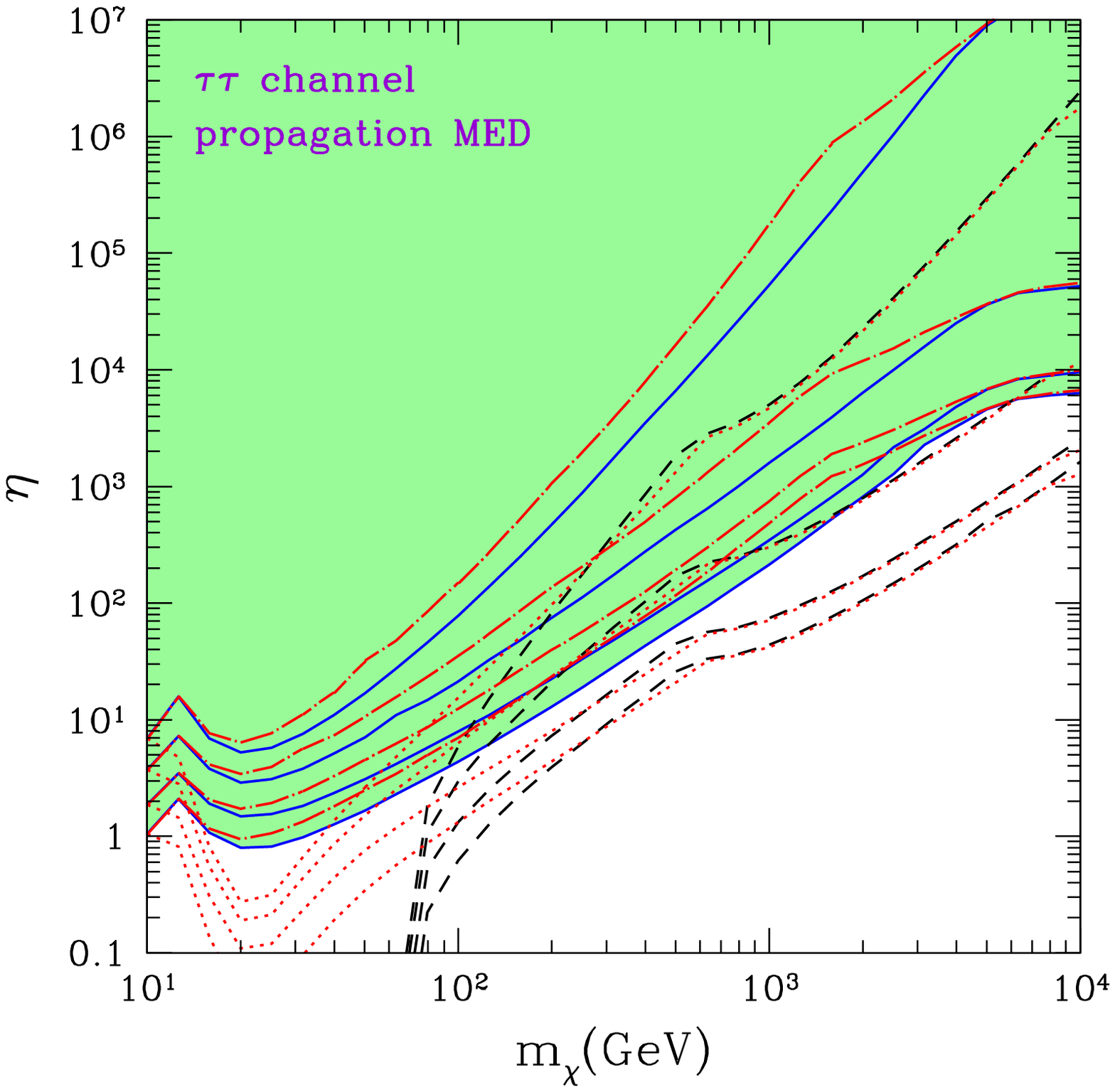}
\caption{The same as in Fig.~\ref{fig:BOUND_eta_mchi_ee.ps}, for DM annihilation into $\tau^+\tau^-$.}
\label{fig:BOUND_eta_mchi_tautau.ps}
\end{figure}

\begin{figure}[t]
\includegraphics[width=1.1\columnwidth]{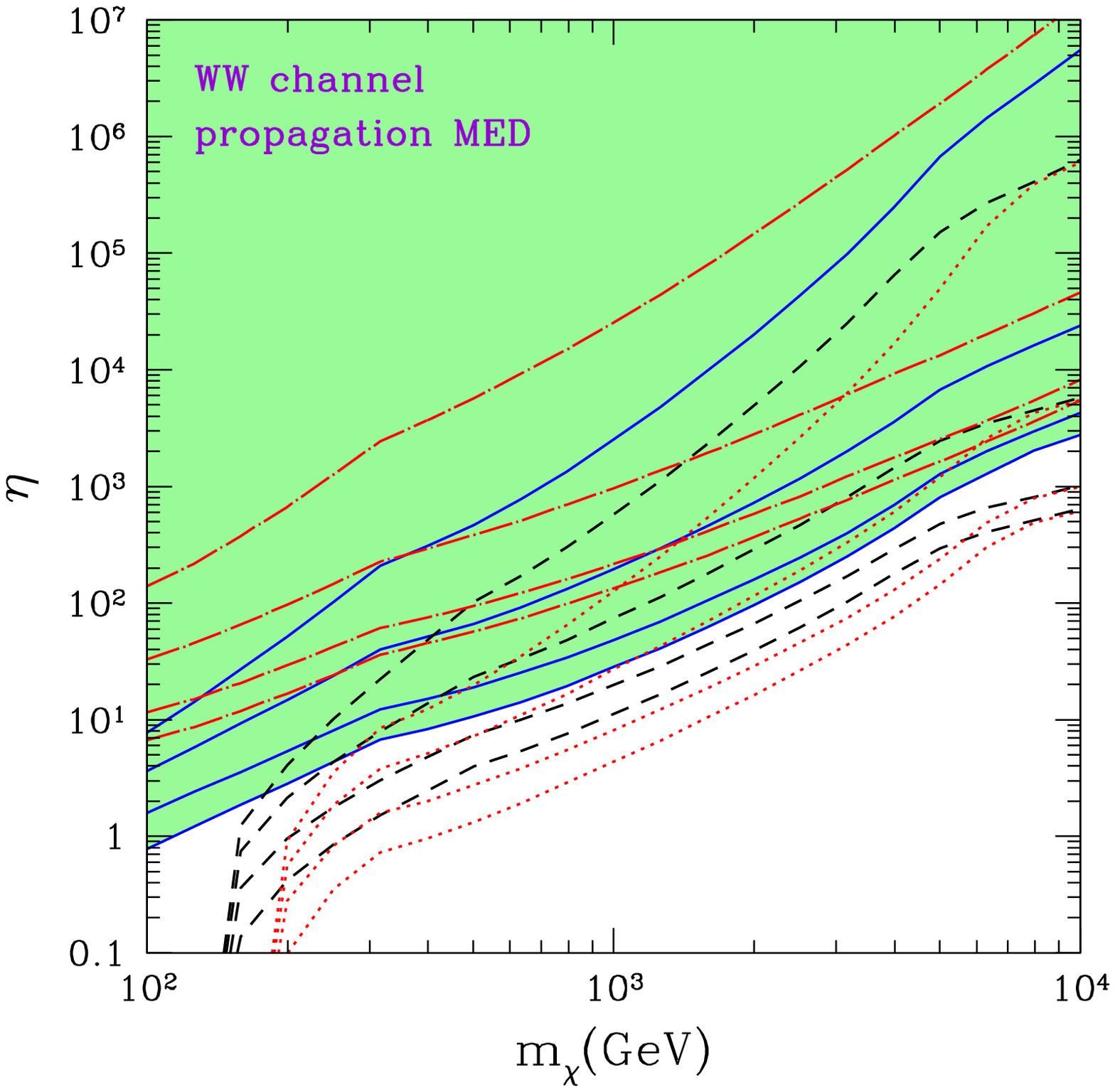}
\caption{The same as in Fig.~\ref{fig:BOUND_eta_mchi_ee.ps}, for DM annihilation into $W^+W^-$.}
\label{fig:BOUND_eta_mchi_WW.ps}
\end{figure}

\begin{figure}[t]
\includegraphics[width=1.1\columnwidth]{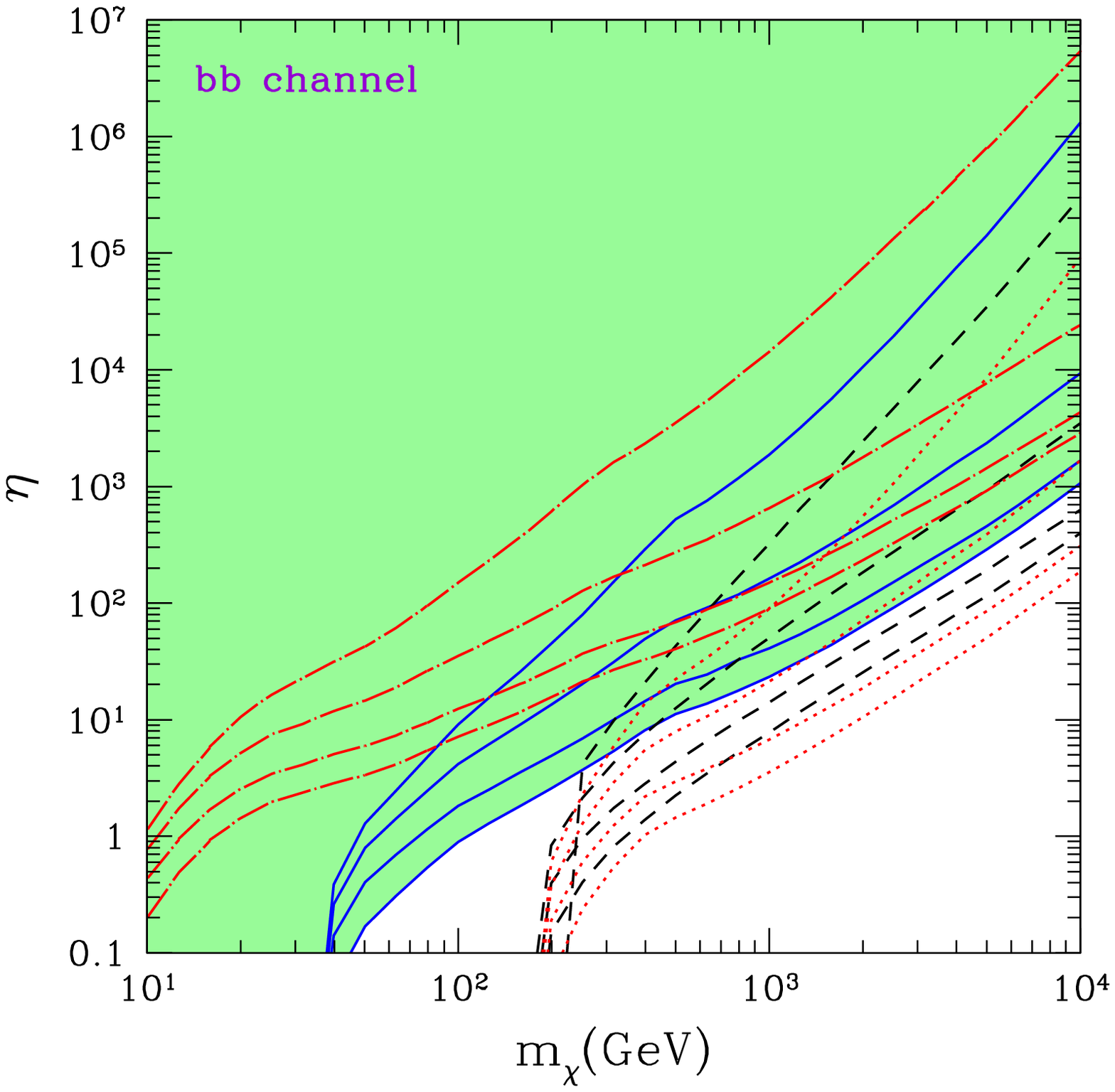}
\caption{The same as in Fig.~\ref{fig:BOUND_eta_mchi_ee.ps}, for DM annihilation into $\bar{b}b$.}
\label{fig:BOUND_eta_mchi_bb.ps}
\end{figure}

%%%%%%%%%%%%%%%%%%%%%%%%%%%%%%%%%%%%%%%%%%%%%%%%%%%%%%%%%%%%%%%%%%%%%%%%%

We start our analysis by studying under what conditions the modified cosmological
scenarios can explain the PAMELA data without violating any of the constraints described in Section \ref{sec:survey}.
The study is performed as follows: we numerically solve the Boltzmann equation for
the evolution of the number density of a thermal relic in a cosmology where the
Hubble rate is given by Eqs.~(\ref{eq:modhubble},\ref{eq:Afunction}) and we determine
the values of the enhancement parameter $\eta$ which are required, for any given cosmology
labelled by the parameter $\nu$, in order to reproduce the correct value $\Omega_{\chi} h^{2} = 
(\Omega h^{2})^{\rm WMAP}_{\rm CDM} = 0.11$
for the relic abundance of the dark matter particle. The annihilation cross sections
are fixed to explain the PAMELA data (within the implemented astrophysical bounds), and have been derived in
Section \ref{sec:survey}. The analysis is performed separately for each of the possibile DM annihilation channels. Analytic considerations may be found in Ref.~\cite{Schelke:2006eg}.

First of all, we show the effect of the modified expansion rate on the dark matter decoupling: this
will be useful to understand the following analysis. Fig.~\ref{fig:decoupling.ps} shows the temperature
evolution of the comoving abundance $Y=n/s$, where $n$ is the number density and $s$ the total entropy density.
The almost--vertical line refers to the Boltzmann--suppressed equilibrium abundance. The other lines refer
to the solution of the Boltzmann equation for various cosmologies, and explicitely show the decoupling 
and asymptotization of $Y$, from which the relic abundance is determined as usual as:
$\Omega_\chi h^2 = m_\chi s_0 Y_0/\rho_c$, where $s_0$ and $Y_0 = Y(T=T_0)$ are the current values of the
entropy density and of $Y$, and $\rho_c$ denotes the critical density of the Universe.
 All the curves are normalized to
the asymptotic value obtained in standard cosmology (GR, for General Relativity), in order to explicitely
show the amount of the enhancement obtained in the different cosmological models. The mass of the
dark matter particle and annihilation cross section have been fixed, in this example, at: $m_\chi = 100$ GeV and $\sigmav = 2.1\cdot 10^{-26}$ cm$^3$ s$^{-1}$. The solid lines refer to cosmologies labelled by the
value of the $\nu$ parameter, enhancement factor $\eta=100$ and re--enetering temperature $T_{\rm re} = 1$ MeV: from top to bottom, $\nu=-1,0,1,2$. We notice that larger values of $\nu$ endow smaller enhancements of the relic abundance,
for the same value of $\eta$: this is due to the fact that a larger $\nu$ implies an Hubble rate which evolves
faster in temperature, and in this case $H(T)$ paces closer to the fast--decreasing annihilation rate, inducing a longer phase of
post--freeze--out annihilation, which in turn reduces the abundance. This fact implies that larger
enhancement factors are required for large values of $\nu$ in order to boost the relic abundace to the WMAP
value, when starting with a very large annihilation cross section, like the case for explaining the PAMELA data.
We will find this behaviour in the following analysis. Fig.~\ref{fig:decoupling.ps} shows also the effect of anticipating the temperature $T_{\rm re}$ at which the modified expansion rate recovers the GR behaviour: dot--dashed
lines refer to $T_{\rm re} = 100$ MeV, while dashed line to $T_{\rm re} = 1$ GeV. We notice, in these cases, that
early re--entering phases are accompanied by a reduction in the relic abudance after $T_{\rm re}$. This phenomenon was explained in Ref.~\cite{Catena:2004ba}, where it was named ``re--annihilation": the (sharp) drop in the enhanced
Hubble rate at $T_{\rm re}$ may result in a new efficient phase of annihilation of the relic particles, because
their annihilation rate returns to be larger than the expansion rate for some time. This effect is clearly
manifest in Fig.~\ref{fig:decoupling.ps}: in this case, larger enhancement parameters $\eta$ would be
required in order to boost enough the relic abundance to the WMAP value, when $\sigmav$ is large. We
remind again that we will always use $T_{\rm re} = 1$ MeV, unless explicitely stated. Finally, we wish to
comment that also
the speed at which the enhanced Hubble rate recovers GR may modify the final relic abundance. We modelled this
phase with the $\tanh$ behaviour of Eq.~(\ref{eq:Afunction}): a change in this behaviour may
(relatively slightly) change the results shown in Fig.~\ref{fig:decoupling.ps}. For definiteness, we will
use the form of Eq.~(\ref{eq:Afunction}) throughout the Paper.

\subsection{Boosts for PAMELA}

Let us now turn to the discussion of cosmological models featuring a thermal relic able to reproduce the right amount of positrons to explain the PAMELA data without being at odds with other astrophysical data, and presenting the correct relic abundance. This is an
alternative solution to the PAMELA ``anomaly" in terms of dark matter annihilation: the compatibility
between the large annihilation cross--sections required by the PAMELA data and the WMAP value of
the relic abundance are obtained by means of modified cosmologies.

Fig.~\ref{fig:PAMELA_eta_mchi_ee.ps} refers to the case of annihilation into an $e^+e^-$
final state, and the annihilation cross--sections used to obtain Fig.~\ref{fig:PAMELA_eta_mchi_ee.ps}
are those shown in Figs.~\ref{fig:boundsVL2ee.eps}, \ref{fig:boundsAQUee.eps} and
\ref{fig:boundsISOee.eps}
which refer to the ``$\frac{e^+}{e^++e^-}$ best--fit'' curve restricted to the range where it is not excluded by any of the considered constraints. 
The values of $\eta$ required to explain the PAMELA positron fraction (and compatible with the other bounds) change significantly
with the cosmological scenario: they are confined in the
range between 0.1 and 100 for $\nu=-1$, while
for $\nu=2$ they are significantly larger, being in an interval from ${\cal O}(1)$ to about $10^4$, depending
on the dark matter mass and on the dark matter halo profile. The values of $\eta$ increase with $m_\chi$ because the PAMELA
data require larger annihilation cross sections for larger masses, as seen in 
Fig.~\ref{fig:boundsVL2ee.eps}: in this case, in order to match the WMAP value for
the relic abundance, larger cross sections require earlier decoupling of the dark matter
particle, and this in turn requires a faster expansion of the Universe, hence larger $\eta$.
In addition, from Fig.~\ref{fig:PAMELA_eta_mchi_ee.ps} we notice that
the enhancement factors are significantly larger for larger values of 
$\nu$, in accordance to the previous discussion in relation with Fig.~\ref{fig:decoupling.ps}.
Fig.~\ref{fig:PAMELA_eta_mchi_ee.ps} also shows that the uncertainty arising from different choices
of the dark matter distribution in the halo may be relevant: the solid lines refer to the case of
the Via Lactea II dark matter distribution, the dot--dashed lines to a cored isothermal sphere and the
dotted lines to the Aquarius simulation. The
results shown in Fig.~\ref{fig:PAMELA_eta_mchi_ee.ps} have been obtained for the MED set of astrophysical
parameters governing cosmic--ray diffusion.

The effect induced by the uncertainties on galactic diffusion is shown in Fig.~\ref{fig:PAMELA_eta_mchi_ee_band.ps}: the yellow
band encompasses the variability on the required values of $\eta$, when the propagation parameters
are changed from the MIN (upper line of each set of curves) to the MAX (lower lines) values. The range
of masses differs, when changing the propagation parameters, because the compatibility of the explanation of PAMELA
data with the other bounds on indirect searches pins down different mass intervals, as discussed in
Section \ref{sec:survey}.

Figs.~\ref{fig:PAMELA_eta_mchi_mumu.ps} and \ref{fig:PAMELA_eta_mchi_tautauWW.ps} report the enhancement
factors required in the case of annihilation into $\mu^+\mu^-$, $\tau^+\tau^-$ and $W^+W^-$ final states.
In these cases, the propagation parameters are set at the MED configuration. The solid, dot--dashed and dotted lines
again refer to Via Lactea II, cored isothermal sphere and Aquarius dark matter distributions, respectively. The range
of masses able to explain PAMELA data and compatible with astrophysical observations are obtained by the analysis of Section \ref{sec:survey}: we notice
the significant difference for the $\tau^+\tau^-$ case, where the isothermal distribution is compatible
with the PAMELA data for a mass interval much larger than for the Via Lactea II or Aquarius cases. For the
$W^+W^-$ final state, Via Lactea II provides only marginal compatibility with the PAMELA data, and is therefore
not present in Fig. \ref{fig:PAMELA_eta_mchi_tautauWW.ps}.
Fig.~\ref{fig:PAMELA_eta_mchi_mumu.ps}, for the $\mu^+\mu^-$ channel, shows also the effect of anticipating
the recovering of the GR cosmic evolution: the dotted lines refer here to $T_{\rm re} = 100$ MeV, for the
isothermal dark matter distribution. As discussed above,
in this case the predicted relic abundance is lower than for the case of lower values of $T_{\rm re}$: as
a consequence, a larger enhancement $\eta$ is necessary to match the correct value of the relic abudance.
This effect similarly affects all the other results shown in Figs.~\ref{fig:PAMELA_eta_mchi_ee.ps},
\ref{fig:PAMELA_eta_mchi_ee_band.ps} and \ref{fig:PAMELA_eta_mchi_tautauWW.ps} and it will affect also similarly
the results on the bounds, to which we now turn.

\subsection{Astrophysical bounds on modified cosmologies}

The astrophysical bounds on the dark matter annihilation cross--section discussed in Section \ref{sec:survey}
may be alternatively used to set constraints on the cosmological histories, as it was done in 
Refs.~\cite{Schelke:2006eg,Donato:2006af}, where we used antiproton and gamma--ray data and a dark
matter particle annihilating dominantly into a quark--antiquark final state (namely $\bar{b}b$).
We now extend that analyses by considering
the whole host of experimental data of Section \ref{sec:survey} and by including the whole set of annihilation
final states of a cold dark matter particle.

The results are shown in Fig.~\ref{fig:BOUND_eta_mchi_ee.ps} for the $e^+e^-$ annihilation channel,
in Fig.~\ref{fig:BOUND_eta_mchi_mumu.ps} for the $\mu^+\mu^-$ channel, in Fig.~\ref{fig:BOUND_eta_mchi_tautau.ps}
for the $\tau^+\tau^-$ channel, in Fig.~\ref{fig:BOUND_eta_mchi_WW.ps} for $W^+W^-$ and in 
Fig.~\ref{fig:BOUND_eta_mchi_bb.ps} for the annihilation into a $\bar{b}b$ pair. In every figure, the solid
lines refer to a cored dark matter distribution, the dashed lines to the Via Lactea II numerical
simulation and the dotted lines to the Aquarius DM distribution. Propagation parameters are set at the MED configuration. The bound for each cosmology 
($\nu=-1,0,1,2,$ from bottom to top) is the area above the corresponding line. We notice that, depending
on the dark matter mass and on the annihilation channel, the bounds may be quite restrictive. This is an
interesting result, since it imposes strong bounds on the cosmological histories of the Universe at the
time of dark matter freeze--out (from $T \sim 400$ MeV to $T \sim 400$ GeV for the mass range considered here), 
under the hypothesis that dark matter is a thermal relic. 
The bounds are
typically stronger for lighter dark matter, since for lighter dark matter the astrophysical bounds are
stronger. The dependence on the dark matter profile may be large and the size of the difference depends also
on the annihilation final state of the DM particle. The same (a dependence on the final annihilation state)
also occurs in determining whether a bound is stronger for Via Lactea II or for Aquarius (the cored isothermal
halo being always less constraining): this occurs because the absolute bound on the DM annihilation cross
section has origin from different signals for different masses and the impact of the DM halo profile affects
differently the various signals, as can be seen for the analysis of Section \ref{sec:survey}. 

The effect of the propagation parameters of charged cosmic--rays is also shown in 
Figs.~\ref{fig:BOUND_eta_mchi_ee.ps} to \ref{fig:BOUND_eta_mchi_bb.ps}, where the dot-dashed lines refer to the MIN configuration. 
We notice that for the MIN case the
bounds are much looser in Figs.~\ref{fig:BOUND_eta_mchi_WW.ps} and \ref{fig:BOUND_eta_mchi_bb.ps}: this is easily understandable from the fact that these are hadronic annihilation
channels and, especially for light dark matter, the bound comes from antiproton searches. The propagation
parameters induce a large uncertainty on the antiproton flux \cite{Donato:2003xg}, and the MIN configuration predicts almost
an order of magnitude smaller flux than the MED case: this implies that compatibility of the antiproton
flux with the data allows larger annihilation cross--sections, and therefore looser bounds on $\eta$ are
obtained.

%%%%%%%%%%%%%%%%%%%%%%%%%%%%%%%%%%%%%%%%%%%%%%%%%%%%%%%%%%%%%%%%%%%%%%%%%
\section{A simple model}\label{secModel}

\begin{figure}[t] 
\centering
\includegraphics[width=1.0\columnwidth]{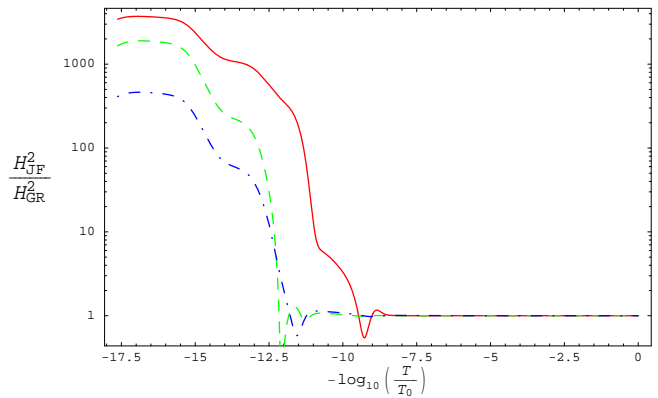}
\caption{Evolution of the (squared) enhancememt factor of the Hubble rate in our modified cosmology,
as a function of the temperature of the Universe. The solid line corresponds to the scalar field parameters $\vp_{\rm in} = 1.9$, $\vp^{\prime}_{\rm in} = 0.45$ and $b=8$ (Model 1); the dashed line to $\vp_{\rm in} = 1.5$, $\vp^{\prime}_{\rm in} = 0.4$ and $b=8$ (Model 2); the dot--dashed line stands for $\vp_{\rm in} = 1.5$, $\vp^{\prime}_{\rm in} = 0.4$ and $b=4$ (Model 3).} 
\label{fig:ratio.eps}
\end{figure}

\begin{figure}[t] 
\centering
\includegraphics[width=1.0\columnwidth]{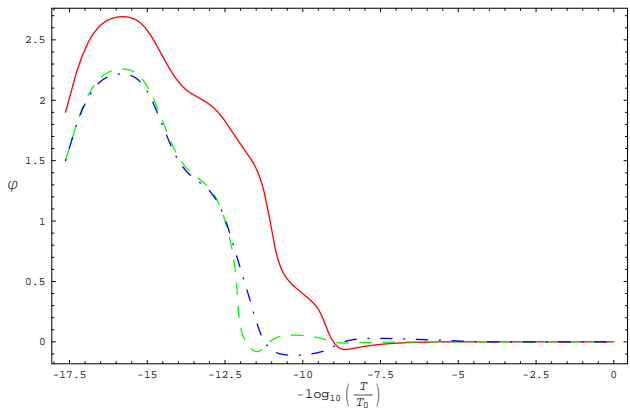}
\caption{Evolution of the scalar field as a function of the temperature of the Universe. The solid line corresponds to $\vp_{\rm in} = 1.9$, $\vp^{\prime}_{\rm in} = 0.45$ and $b=8$ (Model 1); the dashed line to $\vp_{\rm in} = 1.5$, $\vp^{\prime}_{\rm in} = 0.4$ and $b=8$ (Model 2); the dot--dashed line to $\vp_{\rm in} = 1.5$, $\vp^{\prime}_{\rm in} = 0.4$ and $b=4$ (Model 3).}
\label{fig:field.eps}
\end{figure}

\begin{figure}[t] 
\centering
\includegraphics[width=1.0\columnwidth]{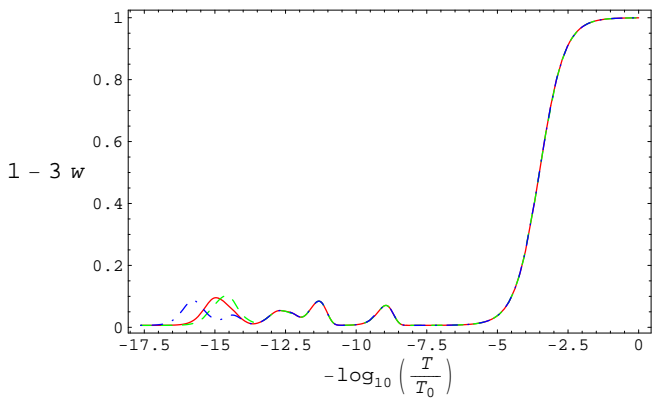}
\caption{Factor $(1-3w)$ as a function of the temperature of the Universe. Notice the fluctuations associated to the different particles becoming non--relativistic in the thermal bath. In this plot we considered a ``SUSY--like'' spectrum. The dashed, solid and dot--dashed lines refer to DM candidates with a mass of 50 GeV, 200 GeV and 1 TeV, respectively.} 
\label{fig:EOStot.eps}
\end{figure}

\begin{figure}[t]
\centering
\includegraphics[width=1.1\columnwidth]{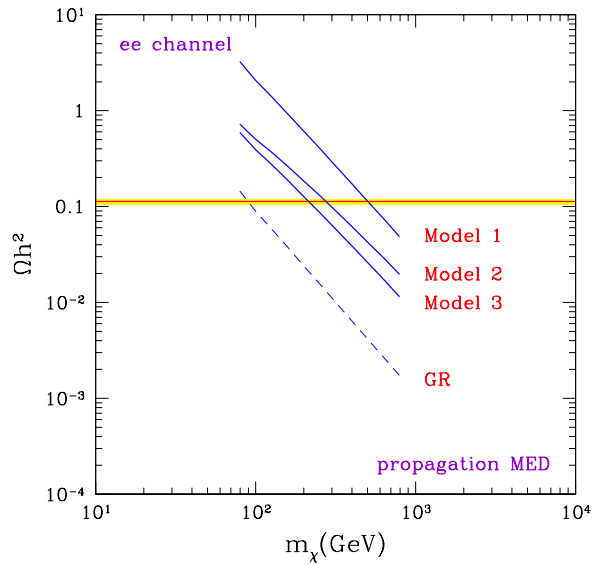}
\caption{Relic abundance as a function of the DM mass, calculated for the three cosmological models of
Fig.~\ref{fig:ratio.eps} and for a DM annihilation cross section able to explain the PAMELA data within the astrophysical constraints for annihilation
into $e^+e^-$, as shown in Fig.~\ref{fig:boundsVL2ee.eps}. The dashed line shows the correspoding relic abundance
in standard cosmology. The DM distribution is from Via Lactea II and the cosmic--ray propagation parameters are set at the
MED case. The horizontal thin band shows the WMAP range for CDM abundance, $(\Omega h^2)_{CDM}^{WMAP}=0.1131 \pm 0.0034$ \cite{WMAP5y}.}
\label{fig:COSMO_omega_mchi_ee.ps}
\end{figure}

\begin{figure}[t]
\centering
\includegraphics[width=1.1\columnwidth]{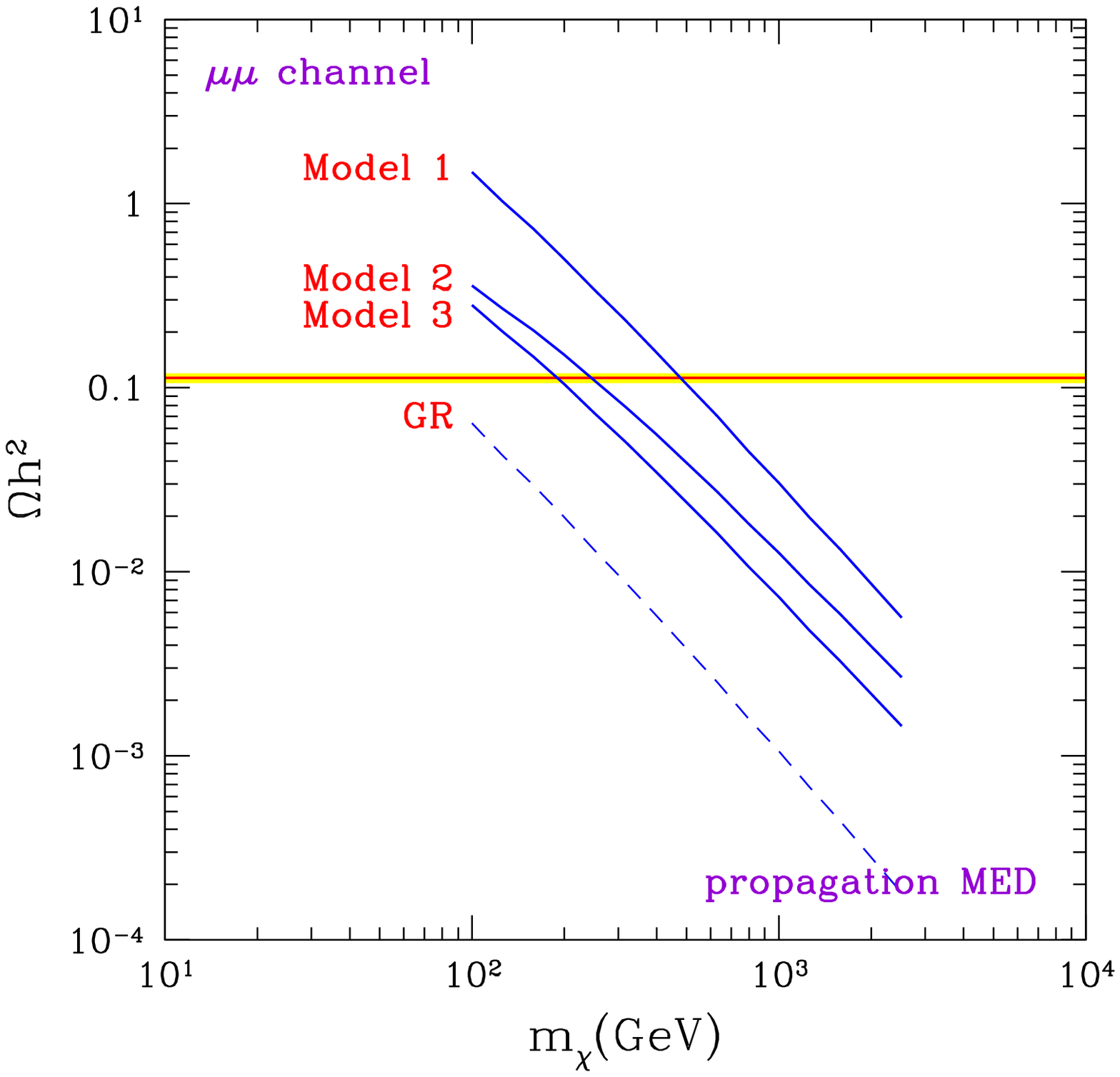}
\caption{The same as in Fig.~\ref{fig:COSMO_omega_mchi_ee.ps}, for DM annihilation into $\mu^+\mu^-$.}
\label{fig:COSMO_omega_mchi_mm.ps}
\end{figure}

\begin{figure}[t]
\centering
\includegraphics[width=1.1\columnwidth]{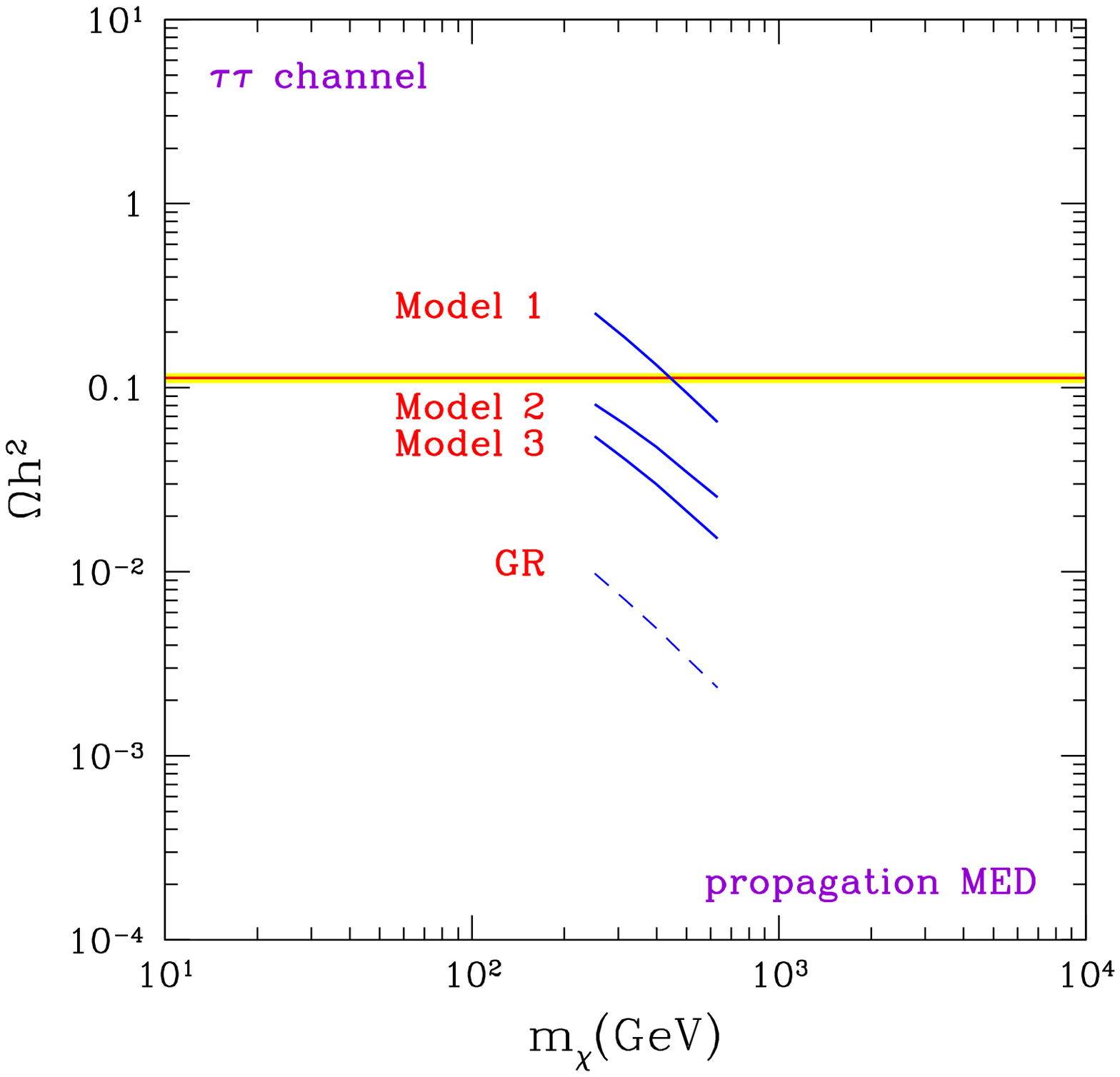}
\caption{The same as in Fig.~\ref{fig:COSMO_omega_mchi_ee.ps}, for DM annihilation into $\tau^+\tau^-$.}
\label{fig:COSMO_omega_mchi_tautau.ps}
\end{figure}

\begin{figure}[t]
\centering
\includegraphics[width=1.1\columnwidth]{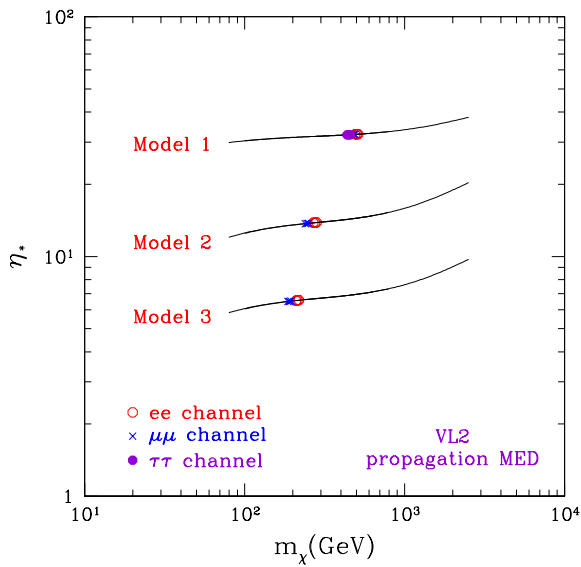}
\caption{Enhancement factor $\eta^*$ of the Hubble rate at the DM decoupling, for the three cosmological models of
Fig.~\ref{fig:ratio.eps}. The open circles, crosses and dots show the configurations which explain the PAMELA data without overproducing other signals in terms of a thermal relic with the correct relic adundance and annihilation into $e^+e^-$, $\mu^+\mu^-$ and $\tau^+\tau^-$, respectively. The DM distribution is from Via Lactea II and the cosmic--ray propagation parameters are set at the
MED case.}
\label{fig:COSMO_eta_mchi.ps}
\end{figure}

\begin{figure}[t]
\centering
\includegraphics[width=1.1\columnwidth]{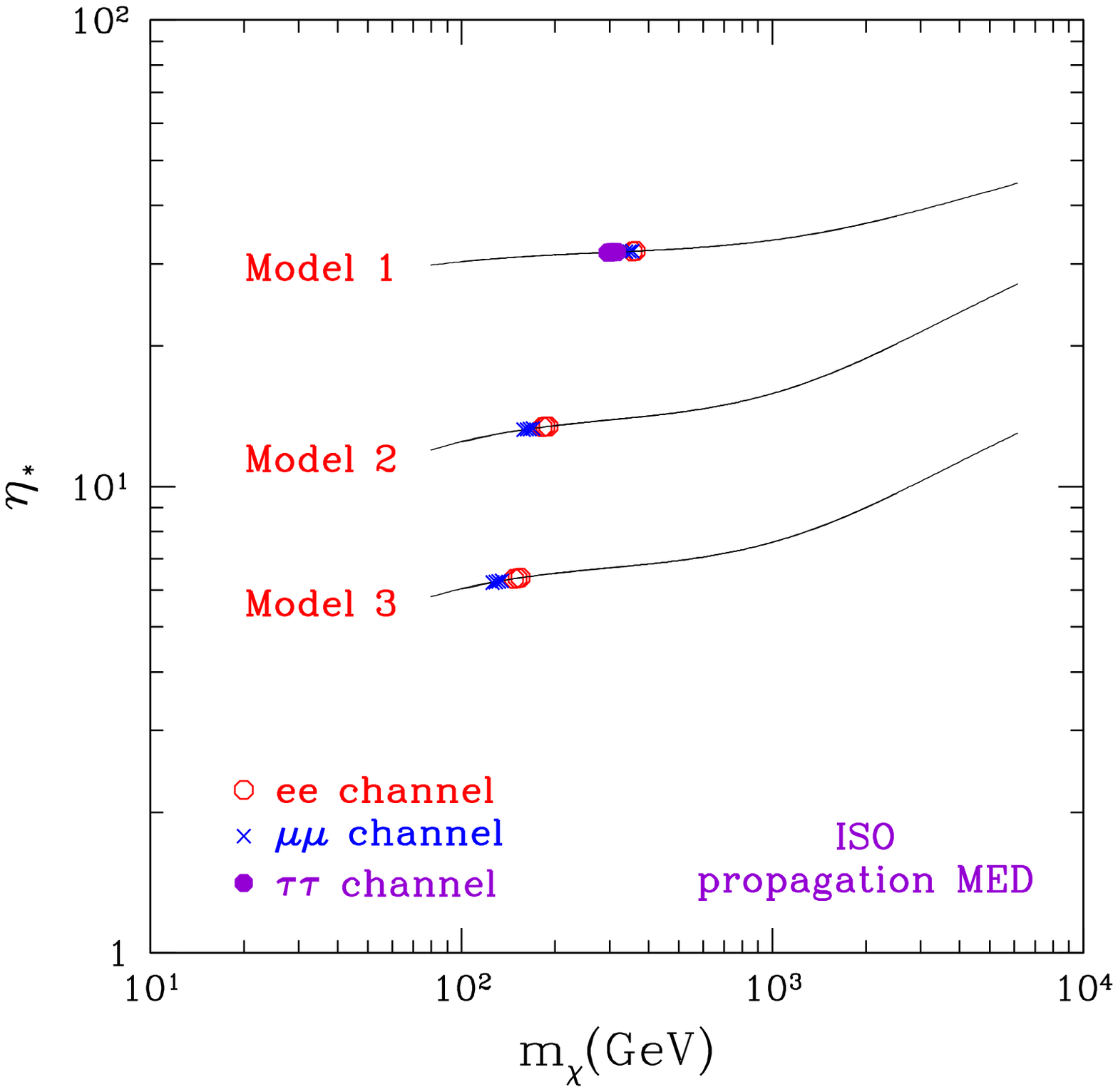}
\caption{The same as in Fig. \ref{fig:COSMO_eta_mchi.ps}, for the cored isothermal halo profile.}
\label{fig:COSMO_eta_mchi_ISO.ps}
\end{figure}

We mentioned in the previous Section that alternative cosmologies, based on extensions of General Relativity, 
can provide the large boost factors required to explain in terms of a thermal relic the positron fraction measured by PAMELA. We now focus on a simple model which allows an explicit calculation of such cosmological boosts.

We consider the action:
\begin{equation}
S=S_{g}+S_{m}\,,  \label{action}
\end{equation}
where:
\begin{equation}
S_{g}=\frac{M_{\rm Pl}^2}{2}\int d^{4}x\sqrt{-{g}}\left\{ {R[g_{\mu \nu}]}+{g}^{\mu \nu }\partial _{\mu }\varphi \partial _{\nu }\varphi -\frac{2}{M_{\rm Pl}^2} V(\varphi )\right\} 
\end{equation}
describes the gravitational interaction, which is now mediated by the metric $g_{\mu \nu}$ and a scalar field $\varphi$. $S_m$ represents the action for the (visible and dark) matter sector. We assume an universal metric coupling between gravity and matter, that is:
\begin{equation} 
S_{m} = S_{m}[\Psi_{m},\A^{2}(\varphi ){g}_{\mu \nu }] \,\,\, , 
\label{sm}
\end{equation}
with $\Psi_{m}$ indicating a generic field of the matter sector coupled to the metric $\A^{2}(\varphi ){g}_{\mu \nu }$. In the present discussion the potential $V(\varphi)$ plays no r\^ole and we therefore set it to zero. Thus, the function $\A(\varphi)$ univocally specifies the model.

If $\A(\varphi)$ is constant, then  the action $S_g + S_m$ is just that of General Relativity, plus a minimally coupled scalar field.  The scalar contribution to the gravitational interaction is therefore measured by the quantity:
\beq 
\alpha(\varphi) = \frac{d \log \A(\varphi)}{d \varphi}\,.
\label{alphab}
\eeq
If $\alpha(\varphi) \ne 0$, then Eq.~(\ref{action}) defines a Scalar-Tensor theory of gravity formulated in the Einstein frame \cite{Damour1}. 

We now consider an homogeneous cosmological space-time: 
\beq
ds^{2}= dt^{2} - a^{2}(t)\ dl^{2}\ , 
\eeq
where the matter energy--momentum tensor, $T_{ \mu \nu } = 
2(-g)^{-1/2}   \delta S_{m} / \delta g^{\mu\nu}$  admits the perfect-fluid representation: 
\beq 
T_{ \mu \nu } 
=(\rho +p)\ u_{\mu }u_{\nu }\ - p\ g_{\mu \nu }\ \ , 
\eeq
with $g_{\mu \nu }\ u^{\mu }u^{\nu }=1$.

The Friedmann-Robertson-Walker (FRW) equations then take the form:
\begin{eqnarray}
&& \frac{\ddot{a}}{a} = - \frac{1}{6 \Mp^2} \left[ \rho +3\ p +2 \Mp^2\dot\varphi^{2} \right]\nonumber\\
&&\left( \frac{\dot{a}}{a}\right) ^{2}+\frac{k}{a^{2}} = \frac{1}{3 \Mp^2}
\left[\rho + \frac{\Mp^2}{2}\dot\varphi^{2} \right] \nonumber\\
&&\ddot{\varphi} +3\frac{\dot{a}}{a}\dot\varphi = 
-\frac{1}{\Mp^2}\left[ \alpha (\rho-3p) 
 \right] 
 \label{eqback}
\end{eqnarray}
where $p=w\, \rho$ and the Bianchi identity reads:
\begin{equation}
{}d(\rho \,a^{3})+p\ da^{3}=(\rho -3\ p)\ a^{3}d\log \A(\varphi ). 
\label{bianchi}
\end{equation}

In the previous Sections we calculated the particle physics rates and cross sections in the Jordan frame, which is defined by the Weyl rescaling $g^{\rm JF}_{\mu \nu} = \A^{2}(\varphi ){g}_{\mu \nu }$ (see Eq.~(\ref{sm})) \cite{old3}. For consistency, we thus study the evolution of the Universe in the Jordan frame. The new degree of freedom $\varphi$ modifies the General Relativity Hubble expansion as follows:
\beq
\frac{H^2_{\rm JF}}{H^2_{\rm GR}} = \A^2(\vp ) \, 
\frac{\left( 1 + \alpha (\vp )\,\vp ^{\prime} \right) ^2}{1-(\vp ^{\prime})^2 /6} \, \equiv A^2(T)\, ,
\label{ratio}
\eeq 
where $H_{\rm GR}$ is the General Relativity Hubble rate and a prime denotes a derivative with respect to the logarithm of the scale factor.  

We now focus on the time (or temperature) evolution of Eq.~(\ref{ratio}) in the specific model defined by $\A = 1 + b \varphi^2$, where $b$ is a constant. This structure follows, for instance, by the assumption of invariance with respect to a discrete $Z_{2}$ symmetry under which the field $\varphi$ is odd. The model has three free parameters: the initial scalar field configuration $(\varphi_{\rm in},\varphi^{\prime}_{\rm in})$ and the constant $b$. 

It is well known that the system (\ref{eqback}) has General Relativity as a late time attractor solution \cite{Damour2}. Moreover, it has  been found that, under very general assumptions, any departure from the GR attractor is associated to an enhancement of the Hubble rate \cite{Catena:2004ba,old2}. The strongest constraint on the model comes from BBN which states that the Hubble rate at the BBN can at most differ by $10\%$ from its GR predicted value \cite{Lisi:1999ng,Olive:1998vj,Cyburt:2004yc,Cyburt:2004cq,Barger:2003zg,Cuoco:2003cu,Mangano:2006ur,Hansen:2001hi}. We found, however, that the GR attractor is so efficient that even regions of the parameter space leading to pre--BBN values of the ratio (\ref{ratio}) order of magnitudes larger than one are allowed by the BBN bound. 
Fig.~\ref{fig:ratio.eps} shows the evolution of the ratio (\ref{ratio}) for three benchmark points in the parameter space of our reference model: the solid line corresponds to the scalar field parameters $\vp_{\rm in} = 1.9$, $\vp^{\prime}_{\rm in} = 0.45$ and $b=8$ (Model 1); the dashed line to $\vp_{\rm in} = 1.5$, $\vp^{\prime}_{\rm in} = 0.4$ and $b=8$ (Model 2); the dot--dashed line stands for $\vp_{\rm in} = 1.5$, $\vp^{\prime}_{\rm in} = 0.4$ and $b=4$ (Model 3).

The efficiency of the attractor is related to the evolution of the scalar field during radiation domination: 
throughout this epoch, when the temperature of the Universe crosses the value corresponding to the mass of a particle in equilibrium in the thermal bath, the right hand side of the field equation becomes different from zero. This makes the field evolve towards its value at the GR attractor. Such a mechanism produces typical features in the field evolution (Fig.~\ref{fig:field.eps}) and in the effective equation of state parameter (Fig.~\ref{fig:EOStot.eps}). 

For the scalar--tensor models introduced so far, we have then calculated the relic abundance for thermal relics
able to explain the PAMELA data and compatible with the other astrophysical constraints, {\em i.e.}~for the annihilation cross sections derived in Section 
\ref{sec:survey}. Figs.~\ref{fig:COSMO_omega_mchi_ee.ps}, \ref{fig:COSMO_omega_mchi_mm.ps} and 
\ref{fig:COSMO_omega_mchi_tautau.ps} show the cases for pure leptonic annihilation channels: $e^+e^-$, $\mu^+\mu^-$
and $\tau^+\tau^-$ respectively. The solid lines refer to Model 1, 2 and 3 introduced above and the dashed line
shows the values of the relic abundance obtained in GR for the values of annihilation cross sections which are
able to explain the PAMELA data within the astrophysical bounds. While in standard cosmology the relic abundance is typically too low
(whence the need for a ``boost"), for
the three scalar--tensor comologies the relic abundance is much larger and may be compatible with the WMAP value,
represented by the horizontal line. We see that in the case of the specific Models 1, 2 and 3 presented here,
solutions to the PAMELA ``anomaly" are found for specific values of the dark matter mass (in a range bewteen
100 and 500 GeV), dependent on the annihilation channel. These specific models therefore explicitly implement
a cosmological solution to the PAMELA ``excess" in terms of a thermal dark matter particle without violating the observations discussed in Section \ref{sec:survey}.

Finally Fig.~\ref{fig:COSMO_eta_mchi.ps} and \ref{fig:COSMO_eta_mchi_ISO.ps} show, in terms of the terminology of the previous Section on
a generic deviation of the expansion rate from the GR behaviour, {\em i.e.}~in terms of the enhancement
parameter $\eta$, the increase produced in Models 1, 2 and 3 here under discussion. Since,
as it was discussed in Section \ref{sec:analysis}, $\eta$ represents in our definition the value of the enhancement
of the Hubble rate at a temperature corresponding to the freeze--out temperature $T^{\rm GR}_{\rm FO}$ in 
GR, we have defined $\eta^*$ shown in Fig.~\ref{fig:COSMO_eta_mchi.ps} as: 
\begin{equation}
1+ \eta^* = A(T^{\rm GR}_{\rm FO}) = \frac{H_{\rm JF}}{H_{\rm GR}}(T^{\rm GR}_{\rm FO})
\end{equation}
We see that Models 1, 2 and 3 endow with enhancements of the order of $5\div50$. The open circles, crosses
and filled circles in Fig.~\ref{fig:COSMO_eta_mchi.ps} (Via Lactea II) and Fig.~\ref{fig:COSMO_eta_mchi_ISO.ps} (cored isothermal) show the values of dark matter masses for which accomodate PAMELA positron data while being compatible with the other bounds is possible, for the $e^+e^-$, $\mu^+\mu^-$ and $\tau^+\tau^-$ channels, respectively.
We can see that for a cored isothermal halo profile, the cosmological model require slightely lighter dark matter in order
to explain the PAMELA data, as compared to the Via Lactea II halo.

\section{Conclusions}
\label{sec:conclusions}

Alternative cosmologies, based on extensions of General Relativity, predict modified thermal histories
in the Early Universe in the pre--BBN era, which is not directly constrained by cosmological observations.
A typical prediction is that the expansion rate is enhanced with respect to the GR case: this, in turn,
implies that thermal relics typically decouple earlier and with larger relic abundances. The correct
value of the relic abundance for a thermal relic, value which cannot exceed the cosmological
determination of the dark matter content of the Universe, is therefore obtained for larger 
annihilation cross--sections, as compared to standard cosmology. 
%The so--called ``WIMP--miracle" persists also in these alternative cosmologies, but is obtained for larger values of $\sigmav$.

Indirect detection rates of dark matter directly depend on the current value of $\sigmav$ in the galactic
halo. In the case of a dominant s--wave annihilation, which is typical for most of the cold dark matter candidates
in large portions of the parameter space of New Physics models, larger values of $\sigmav$ required to match
the WMAP dark matter abundance in modified cosmologies imply larger signals in the Galaxy. We have
exploited this feature in a twofold way.

First of all, we can use the large host of independent results on the search for indirect signals of dark matter
to set bounds on the enhancement of the Hubble rate in the pre--BBN era: this idea was introduced in
Ref.~\cite{Schelke:2006eg}, and there pursued by using exotic antiproton searches in cosmic--rays. A first
attempt to use the gamma--ray signal was done in Ref.~\cite{Donato:2006af}. In the current Paper we extend
these analyses by introducing a whole set of indirect detection signals, which became increasingly relevant
in the last months with the recent results from detectors like PAMELA, Fermi-LAT and HESS.
We have classified cathegories of cosmological models with an enhanced expansion rate and we have
derived bounds on them under the hypothesis that the dark matter is a thermal relic. The observational
data we have used to set bounds on the dark matter annihilation cross section (from which, in turn,
we have derived the bounds on the cosmological models) come from radio observations, gamma--ray observations,
inverse Compton photons, antiprotons and positrons in cosmic--rays and the optical depth of the Cosmic Microwave Background photons.
The bounds have been derived for a generic dark matter particle, by studying separately the different annihilation
channels. 

The recent results on the measurement of the positron fraction realized by the PAMELA detector, have
shown a clear and steady raise at energies above 10 GeV. This behaviour, which is currently under deep scrutiny,
has one of its interpretations in terms of a dark matter signal through dark matter annihilation in our Galaxy.
The theoretical analyses which have discussed this possibility show that, in order to explain the PAMELA ``excess",
the annihilation cross sections need to be orders of magnitude larger than those required in standard cosmology
to explain the observed amount of cold dark matter. This fact poses a problem for a thermal relic, and 
various mechanisms have been invoked to boost the positron signal without spoiling the correct value of relic abundance
(astrophysical boosts, Sommerfeld enhanced cross--sections). Since in cosmologies with an enhanced expansion rate
we naturally require larger annihilation cross--sections, we have discussed the properties of these alternative
cosmologies in order to be able to explain the PAMELA ``puzzle" without violating other observations. We have derived the required amount of
enhancement for different cosmologies.

Finally, in order to demonstrate the feasibility of such an approach, we have
discussed scalar--tensor theories of gravity, for which we constructed explicit models able to reproduce
the required cosmological boost to explain the PAMELA data while being compatible with astrophysical constraints.

\acknowledgements
Work supported by research grants funded jointly by Ministero dell'Istruzione,
dell'Universit\`a e della Ricerca (MIUR), by Universit\`a di Torino, by
Universit\`a di Padova, by the Istituto Nazionale di Fisica Nucleare (INFN) 
within the {\sl Astroparticle Physics Project}, by the Italian Space Agency (ASI) under 
contract N$^{\circ}$ ASI-INAF I/088/06/0 and by Fundac\~{a}o para a Ci\^encia e Tecnologia 
(Minist\'erio da Ci\^encia, Tecnologia e Ensino Superior). 
This research was also supported in part by the European Programmes 
``Universenet" contract MRTN-CT-2006-035863, ``Unification in the LHC Era" contract 
PITN-GA-2009-237920 (UNILHC) and by the Project of Excellence 
``LHCosmology" of the  Padova Cariparo Foundation. 
NF acknowledges support of the spanish MICINNÕs Consolider--Ingenio 2010 
Programme under grant MULTIDARK CSD2009-00064.
MP would like to thank the International Doctorate on AstroParticle Physics (IDAPP) network. NF warmly thanks the Institut de Physique Th\'eorique (IPhT) of CEA/Saclay and the Institut d'Astrophysique de Paris for the hospitality during the period when this work has been started. LP thanks the Physics Department of the University of Trento for hospitality. Finally, all the authors wish to thank M. Pietroni for long--standing and valuable discussions on many of the topics discussed in this paper.

%%%%%%%%%%%%%%%%%%%%%%%%%%%%%%%%%%%%%%%%%%%%%%%%%%%%%%%%%%%%%%%%%%%%%%%%%%%%%%%%%%%%%%%%

\end{document}